\documentclass[aps,pre,onecolumn,12pt,nofootinbib,longbibliography,floatfix]{revtex4-2}

\usepackage[T1]{fontenc}
\usepackage{newtxtext,newtxmath}
\usepackage{microtype}
\usepackage{amsmath,bm,mathtools}
\usepackage{graphicx}
\usepackage{xcolor}
\usepackage{tikz}
\usetikzlibrary{arrows.meta,calc,positioning}
\usepackage{pgfplots}
\pgfplotsset{compat=1.18}
\usepgfplotslibrary{groupplots,fillbetween}
\usepackage{hyperref}
\usepackage[nameinlink]{cleveref}
\crefname{figure}{Fig.}{Figs.}
\Crefname{figure}{Figure}{Figures}
\crefname{equation}{Eq.}{Eqs.}
\Crefname{equation}{Equation}{Equations}
\crefname{section}{Sec.}{Secs.}
\Crefname{section}{Section}{Sections}
\crefname{subsection}{Sec.}{Secs.}
\Crefname{subsection}{Section}{Sections}
\crefname{appendix}{Appendix}{Appendices}
\Crefname{appendix}{Appendix}{Appendices}

\definecolor{Ink}{HTML}{202B33}
\definecolor{Muted}{HTML}{65737E}
\definecolor{Grid}{HTML}{DCE3E8}
\definecolor{DeepBlue}{HTML}{2E6F9E}
\definecolor{Teal}{HTML}{238574}
\definecolor{Orange}{HTML}{D9822B}
\definecolor{Vermilion}{HTML}{C94C3B}
\definecolor{Purple}{HTML}{76538E}
\definecolor{PaperBlue}{HTML}{0B12C9}
\definecolor{PaperGreen}{HTML}{3CA02C}
\definecolor{PaperBurgundy}{HTML}{650B02}
\definecolor{PaperYellow}{HTML}{FFE94A}
\definecolor{OutsideGray}{HTML}{E4E4E4}
\definecolor{WeakBlue}{HTML}{9CC6DE}
\definecolor{WeakOrange}{HTML}{F3C991}
\definecolor{Burgundy}{HTML}{8C2B27}

\hypersetup{
  colorlinks=true,
  linkcolor=DeepBlue,
  citecolor=DeepBlue,
  urlcolor=DeepBlue,
  pdftitle={Understanding the Mpemba Effect through a Rule of Thumb},
  pdfauthor={Ran Yaacoby}
}

\newcommand{\dd}{\mathrm{d}}
\newcommand{\e}{\mathrm{e}}

\newcommand{\KL}{D_{\mathrm{KL}}}
\newcommand{\Tb}{T_{\mathrm b}}

\newcommand{\Tc}{T_{\mathrm c}}
\newcommand{\Th}{T_{\mathrm h}}

\newcommand{\brk}[1]{\left( #1 \right)}
\newcommand{\bigbrk}[1]{\bigl( #1 \bigr)}
\newcommand{\Brk}[1]{\left[ #1 \right]}
\newcommand{\abs}[1]{\left| #1 \right|}

\newcommand{\avg}[1]{{\left\langle #1 \right\rangle}}

\pgfplotsset{
  paperaxis/.style={
    axis line style={draw=Ink,line width=0.55pt},
    tick style={draw=Ink,line width=0.45pt},
    tick label style={font=\scriptsize,color=Muted},
    label style={font=\footnotesize,color=Ink},
    title style={font=\footnotesize\bfseries,color=Ink,align=left},
    grid=major,
    grid style={draw=Grid,line width=0.3pt},
    legend style={font=\scriptsize,draw=none,fill=white,fill opacity=0.9,text opacity=1},
    every axis plot/.append style={line cap=round,line join=round},
  }
}

\AtBeginDocument{\linespread{1}\selectfont}

\begin{document}

\title{Understanding the Mpemba Effect through a Rule of Thumb}

\author{Ran Yaacoby}
\affiliation{Department of Condensed Matter Physics, Tel Aviv University,
Tel Aviv 69978, Israel}

\begin{abstract}
In the Mpemba effect, a system prepared farther from equilibrium relaxes faster than one prepared closer to it.
Its persistence across diverse systems, and under different dynamics of the same system, suggests that it is largely rooted in equilibrium properties, with the dynamics deciding only which mode relaxes slowest.
Here we propose a simple rule of thumb for anticipating the Mpemba effect: the effect is expected when a single low-order equilibrium moment is nonmonotonic in temperature.
The bath singles out this moment as the slowest to relax, so that its displacement from its bath value sets the amplitude of the slowest relaxation mode.
We establish this relation exactly for low-dimensional Ornstein--Uhlenbeck processes.
We then extend the rule to mean-field models by approximating their relaxation near the final equilibrium state by an Ornstein--Uhlenbeck process.
For the mean-field Ising antiferromagnet, which we solve exactly in the thermodynamic limit, this approximation reproduces the main structure of the Mpemba phase diagram, including strong effects far from any phase transition.
Finally, we give a general prescription for mean-field systems: identify the slowest symmetry-allowed moment of the normal coordinates at the bath, and follow its equilibrium value as a function of preparation temperature.
This offers a practical route from equilibrium observables to anomalous relaxation without diagonalizing the dynamics.
\end{abstract}

\maketitle

When a system is brought into contact with a thermal bath, one usually expects the state that starts closer to equilibrium to be the first to reach it.
The Mpemba effect violates this expectation: a hot system can cool faster than a colder copy of itself, and, in the inverse effect, a cold system can heat faster than a warmer one.
Long known as a puzzle about freezing water \cite{Mpemba1969,Jeng2006,Burridge2016}, it is now a well-defined phenomenon of nonequilibrium statistical physics \cite{Bechhoefer2021,Teza2026review,Ares2025review}.
It has been found in granular gases \cite{Lasanta2017,Torrente2019,Biswas2020}, spin systems \cite{LuRaz2017,Klich2019,BaityJesi2019,Yang2020,Vadakkayil2021,Holtzman2022,Chatterjee2023,GonzalezAdalid2024,Pomares2026}, colloids \cite{Kumar2020,Chetrite2021,Kumar2022,Walker2021,Walker2022mfpt,Biswas2023,Biswas2023pop}, quantum systems \cite{Carollo2021,Ares2023,Joshi2024,Aharony2024,Zhang2025,Campaioli2026} and several others \cite{Schwarzendahl2022,Chaddah2010,Ahn2016,Hu2018}.
It is also related to other anomalous relaxations, such as the Kovacs memory effect \cite{Kovacs1964,Bertin2003,Mossa2004,Bouchbinder2010,RuizGarcia2014,Prados2014,Lahini2017} and the heating--cooling asymmetry \cite{Lapolla2020,*Lapolla2022erratum,Ibanez2024}.
General formalisms describe it through the slowest modes of the relaxation dynamics \cite{LuRaz2017,Klich2019,VanVu2025}, and have been used to design faster relaxation protocols \cite{GalRaz2020,Busiello2021,Boundary2023,Carollo2021,Beato2026}.

A striking feature of these studies is how little the effect depends on the details of the dynamics.
The same system can exhibit the effect when the bath acts on every degree of freedom or only on the boundary \cite{Boundary2023}, when it couples strongly or weakly \cite{Teza2022weak}, under different transition rates with the same equilibrium \cite{Bera2023}, and when inertia is included \cite{AharonyShapira2026}.
This robustness suggests that the effect is rooted in equilibrium properties of the system, and many studies indeed point that way.
The effect has been tied to equilibrium phase transitions \cite{Holtzman2022,Blom2026,Pomares2026}, predicted from equilibrium observables and correlations \cite{Vadakkayil2021,Chatterjee2023}, linked to the temperature dependence of the extractable work \cite{Chetrite2021}, and, in Langevin systems, to the shape of the potential and its confining walls \cite{Kumar2020,Kumar2022,Walker2021,Biswas2023,Biswas2023pop,Liu2026wall,Liu2026predict,Hayakawa2026,Sane2026}.

Yet these links between equilibrium properties and the Mpemba effect remain tied to specific models, and no general principle connects them.
The most general of them approximate the slow mode by a single equilibrium observable, a susceptibility \cite{GonzalezAdalid2024,Pomares2026} or, by symmetry, an order parameter \cite{Blom2026}, near a phase transition.
Which observable should be chosen in general, and when such a choice is justified, remain open.
A second puzzle concerns the bath: in the same system the effect can appear at some bath temperatures and disappear at others, and the boundaries have been associated with changes in the direction along which the system relaxes \cite{Walker2021,Teza2023}, without settling whether these changes control the effect or merely accompany it.
Two questions therefore remain open: which equilibrium properties of the initial state control the effect, and what role the bath plays in selecting them.

\begin{figure}[t]
\centering
\includegraphics[width=\textwidth]{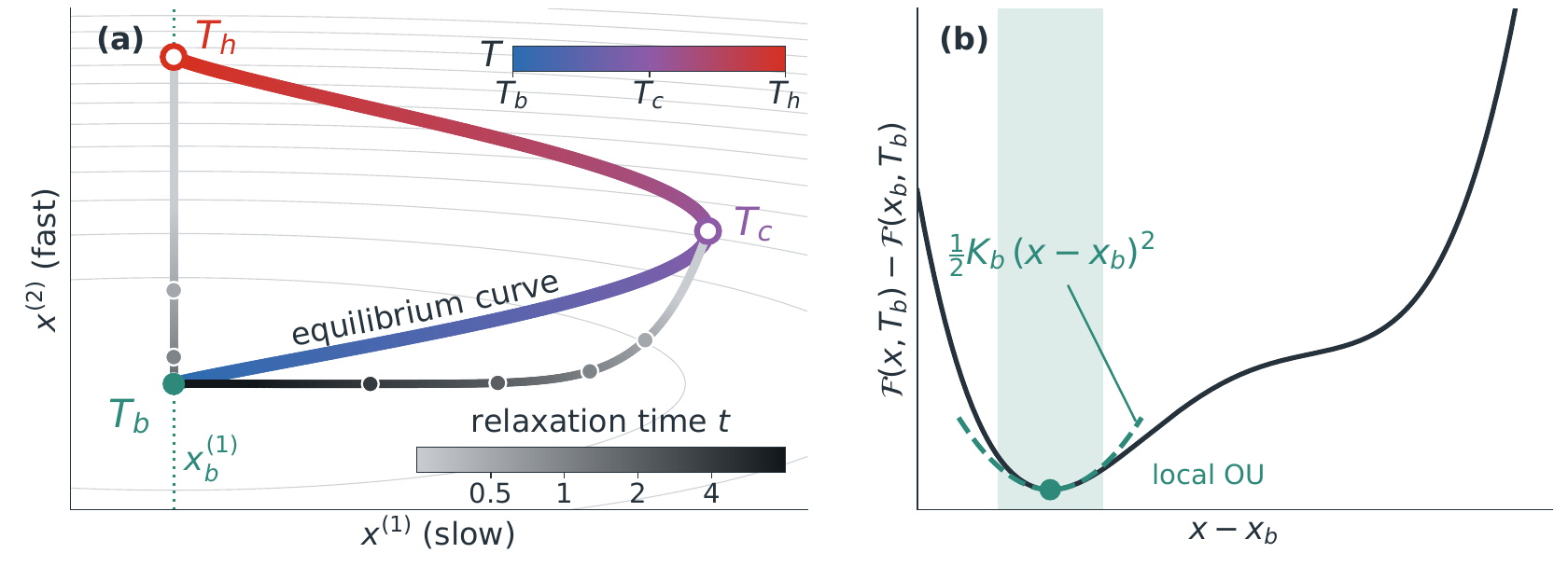}
\caption{Illustration of the mechanism, for a quench from a preparation temperature $T$ to a bath at temperature $\Tb$.
	(a) Top view of a schematic bath free-energy landscape (gray contours) with a slow coordinate $x^{(1)}$ and a fast coordinate $x^{(2)}$.
	The equilibrium curve, colored by $T$, is nonmonotonic in $x^{(1)}$, so the hot preparation $\Th$ starts at the bath value of the slow coordinate, $x^{(1)}_b$ (dotted), and the cold one $\Tc$ does not.
	The relaxation paths are colored by time, with dots at equal times: the hot preparation starts farther away, but it relaxes along the fast coordinate and reaches the bath first.
	(b) Schematic bath free energy (black) and its quadratic approximation near the minimum $x_b$ (dashed), which turns the relaxation into an Ornstein--Uhlenbeck (OU) process (\cref{sec:afm-ou,sec:general}); the shaded band marks this local region.}
\label{fig:mechanism}
\end{figure}

In this work we answer both questions, exactly for simple systems and approximately for systems with many degrees of freedom.
We begin with the exactly solvable Ornstein--Uhlenbeck (OU) process in low dimensions.
We show that the coefficient of its slowest mode is exactly \emph{the displacement of a single equilibrium moment}, such as the mean position of a particle, from its bath value.
A Mpemba effect occurs when this displacement is \emph{nonmonotonic} in the preparation temperature: a hotter preparation then starts farther from equilibrium but closer to the bath value of the moment, and relaxes faster (illustrated in \cref{fig:mechanism}(a)).
Which moment matters is decided by the bath: it is the slowest one allowed by symmetry, and as the bath is varied it can switch from one moment to another, so that the effect depends on whether the newly selected moment is nonmonotonic, not on the switch itself.
We then turn to the mean-field Ising antiferromagnet, which we solve exactly in the thermodynamic limit, and pinpoint the equilibrium moment that controls its relaxation at every bath.
Approximating its dynamics near the bath by an OU process (\cref{fig:mechanism}(b)), which requires only the free energy of its order parameters, reduces it to the same picture: the nonmonotonicity of a single moment predicts the existence of the effect and captures the main structure of the exact Mpemba phase diagram.
This yields a rule of thumb for the Mpemba effect: find the slowest moment allowed by symmetry, and check whether its equilibrium value is nonmonotonic in the preparation temperature, with no need to diagonalize the dynamics.
Finally, we give a prescription, requiring only the free energy and the mobility at the bath, for applying this rule of thumb to general mean-field systems.
In particular, the controlling moment is derived rather than assumed, and need be neither a susceptibility nor tied to a phase transition.

\section{Quantifying the Markovian Mpemba effect}
\label{sec:measuring}

We define the effect following \cite{LuRaz2017,Klich2019}.
In a thermal quench, a system prepared in equilibrium at temperature $T$ is coupled at $t=0$ to a bath at temperature $\Tb$.
Its state is described by the probability $p(x,t)$ to find it in state $x$ at time $t$, starting from the Boltzmann distribution $p(x,0)=\pi(x;T)\propto\e^{-U_T(x)/T}$, where $U_T(x)$ is the potential at temperature $T$.
Throughout, we consider such equilibrium preparations, so each initial state is labeled by a single temperature $T$.
For a continuous overdamped coordinate, $p(x,t)$ evolves under the Fokker--Planck equation, with mobility $\mu$,
\begin{equation}
	 \partial_t p(x,t)=\mathcal L_{\rm FP}\,p=\mu\nabla\!\cdot\!\Brk{p\nabla U_{\Tb}
	 +\Tb\nabla p}.
	 \label{eq:master}
\end{equation}
For a discrete system, $x$ labels the states and \cref{eq:master} is replaced by a master equation with rate matrix $\mathbf W(\Tb)$.
Everything below applies to both cases, with integrals over $x$ replaced by sums; we focus on the continuous case with $\mathcal L_{\rm FP}$.

We expand the relaxation in the eigenfunctions of $\mathcal L_{\rm FP}$,
\begin{align}
	p(x,t)&=\pi(x;\Tb)+\sum_{n\geq1}a_n(T,\Tb)\,\e^{-\lambda_n(\Tb) t}\,v_n(x;\Tb), \label{eq:expansion}\\
	a_n(T,\Tb)&=\int\!\dd x\,u_n(x;\Tb)\,\pi(x;T) \equiv \avg{u_n}_T, \label{eq:a_n}
\end{align}
where $v_n(x;\Tb)$ and $u_n(x;\Tb)$ are the right and left eigenfunctions, and $\lambda_n(\Tb)>0$ are the decay rates, in increasing order.
The rates and eigenfunctions belong to the bath dynamics and therefore depend only on $\Tb$.
The coefficients $a_n(T,\Tb)$ are the overlaps of the initial distribution, $\pi(x;T)$, with the left eigenfunctions $u_n$, that is, the equilibrium averages of $u_n$ at the preparation temperature, and they depend on both temperatures: on the preparation $T$ through $\pi(x;T)$, and on the bath $\Tb$ through $u_n$.
This separation underlies everything that follows: the bath alone decides which mode is slow, and the preparation only sets how strongly it is excited.
For brevity, we omit the $\Tb$ dependence of the eigenfunctions in the rest of the paper, writing $v_n(x)$ and $u_n(x)$.
Every coefficient vanishes when the system is prepared at the bath temperature, $a_n(\Tb,\Tb)=0$.

At late times the slowest mode, $n=1$, dominates, and the distance from equilibrium is set by $\abs{a_1(T,\Tb)}$.
A Mpemba effect occurs when a preparation that starts farther from equilibrium has the smaller slow coefficient,
\begin{equation}
	 \abs{a_1(\Th,\Tb)}<\abs{a_1(\Tc,\Tb)},\qquad \Th>\Tc>\Tb,
	 \label{eq:mpemba-condition}
\end{equation}
for a direct effect, and the analogous condition with $\Th<\Tc<\Tb$ for an inverse effect.
We measure the distance from equilibrium by the Kullback--Leibler (KL) divergence,
\begin{equation}
	\KL(t)=\int\!\dd x\,p(x,t)\ln\frac{p(x,t)}{\pi(x;\Tb)},
	\label{eq:kl}
\end{equation}
with initial value $D_0(T)=\KL(0)$.
The effect is meaningful only if $D_0$ grows monotonically with $\abs{T-\Tb}$: otherwise a hotter preparation may simply start closer to equilibrium, and its faster relaxation is trivial.
In all the examples below $D_0$ is monotonic (\cref{app:kl-monotonic}), so a nonmonotonic $\abs{a_1(T,\Tb)}$ on one side of the bath signals the effect.
It is common to define the strong effect as the case in which $a_1$ has an additional zero, $a_1(T_s,\Tb)=0$ with $T_s\neq\Tb$ \cite{Klich2019}; we call $T_s$ a strong point.
The slowest mode is then absent altogether, and the preparation at $T_s$ relaxes exponentially faster, at the rate $\lambda_2(\Tb)$ instead of $\lambda_1(\Tb)$.
At late times it is then closer to equilibrium than any preparation with $a_1\neq0$ in every distance that decreases monotonically under the dynamics, not only in KL \cite{VanVu2025}.

\section{One dimension: an equilibrium mean controls the slowest mode}
\label{sec:one-dimensional}

\begin{figure*}[t]
	\centering
	\includegraphics[width=\textwidth]{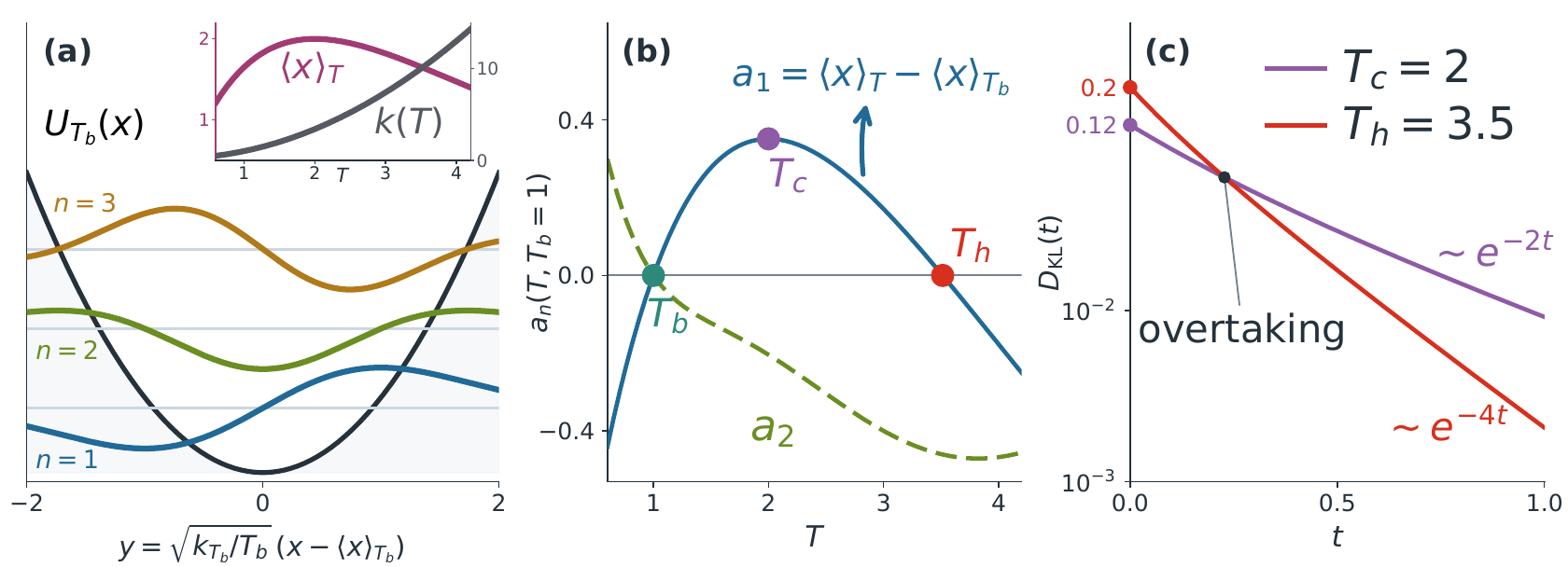}
	\caption{Mpemba effect in the one-dimensional OU process, \cref{eq:bellmodel}, at $\Tb=1$.
		(a) Bath potential in the standardized coordinate $y$ (black), with the first three nonstationary right eigenfunctions $v_n(y)$, \cref{eq:ou1d-spectrum}, for $n=1,2,3$ (blue, green, orange), scaled and offset for visibility.
		Inset: $\avg{x}_T$ (left axis, purple) and $k\brk{T}$ (right axis, gray).
		(b) The two slowest coefficients, \cref{eq:ou1d-coefficients,eq:ou1d-coefficients_2}, against the preparation temperature: $a_1=\avg{x}_T-\avg{x}_{\Tb}$ (solid blue; here $k_{\Tb}=\Tb$) and $a_2$ (dashed green).
		Since $\avg{x}_T$ is nonmonotonic, $a_1$ vanishes not only at the bath but also at $\Th\approx3.5$, and is largest at $\Tc=2$, the maximum of the mean.
		(c) KL divergence, \cref{eq:kl}, after quenches from $\Tc$ and $\Th$: the hot preparation starts farther away but overtakes the cold one, decaying as $\e^{-2\lambda_2t}$ instead of $\e^{-2\lambda_1t}$.}
	\label{fig:bell}
\end{figure*}

We begin with the simplest setting in which the slow coefficient $a_1$ can be computed exactly: a single overdamped coordinate in a harmonic potential whose center $\avg{x}_T$ and stiffness $k \brk{T}$ depend on temperature.
We choose
\begin{equation}
	\begin{aligned}
		 U_T(x)&=\frac{k \brk{T}}{2}\,\Brk{x-\avg{x}_T}^2,\\
		 \avg{x}_T&=T\e^{1-T/2},\qquad
		 k \brk{T}=\frac{1+4T^2}{5}.
	\end{aligned}
	 \label{eq:bellmodel}
\end{equation}
Since $U_T$ is quadratic, the Boltzmann equilibrium state is Gaussian, $\pi(x;T)\propto e^{-\frac{k\brk{T}}{2T}\bigbrk{x-\avg{x}_T}^2}$.
The inset of \cref{fig:bell}(a) shows the temperature dependence of this choice: the mean $\avg{x}_T$ is nonmonotonic, growing up to $T=2$ and decreasing beyond it, while the stiffness $k\brk{T}$ grows monotonically.
With $\mu=1$, the distribution after a quench to $\Tb$ obeys the Fokker--Planck equation of the Ornstein--Uhlenbeck (OU) process \cite{Uhlenbeck1930},
\begin{equation}
	 \partial_tp(x,t)=\partial_x\!\Brk{k_{\Tb}\brk{x-\avg{x}_{\Tb}}p
	 +\Tb\,\partial_xp}.
	 \label{eq:ou1d-fp}
\end{equation}
This system behaves as a thermodynamic one should: the initial KL divergence, $D_0(T)$ in \cref{eq:kl}, grows monotonically as $T$ moves away from $\Tb$ (see \cref{app:kl-monotonic}), so a hotter preparation always starts farther from equilibrium.

For a fixed spring, with $\avg{x}_T=\avg{x}$ and $k\brk{T}=k$ independent of $T$, this process was studied before and shown not to exhibit a Mpemba effect \cite{Walker2021}.
Letting the mean depend on temperature allows us to control the existence of the Mpemba effect exactly, through the first moment. 
The eigenfunctions and decay rates of \cref{eq:ou1d-fp} are known in closed form~\cite{Risken1989,Gardiner2009}; in the standardized coordinate $y$ they read
\begin{equation}
	 v_n(y)=\frac{\operatorname{He}_n(y)\,\e^{-y^2/2}}{\sqrt{2\pi\,n!}},\quad
	 u_n(y)=\frac{\operatorname{He}_n(y)}{\sqrt{n!}},\quad
	 \lambda_n(\Tb)=n\,k_{\Tb}, \quad
	 y\equiv\sqrt{k_{\Tb}/\Tb}\,\brk{x-\avg{x}_{\Tb}}
	 \label{eq:ou1d-spectrum}
\end{equation}
with $n=0,1,2,\ldots$ and Hermite polynomials $\operatorname{He}_n$.
The first right eigenfunctions are drawn in \cref{fig:bell}(a) on top of the bath potential.
Each left eigenfunction is a polynomial of degree $n$, so by \cref{eq:a_n} each coefficient is a combination of equilibrium moments at the preparation temperature.
In particular, the two slowest coefficients, which decay at the rates $\lambda_1(\Tb)=k_{\Tb}$ and $\lambda_2(\Tb)=2k_{\Tb}$, depend only on the first two cumulants of $x$, the mean $\avg{x}_T$ and the variance $T/k \brk{T}$:
\begin{align}
	\operatorname{He}_1(y)=y\quad&\longrightarrow\quad
	a_1(T,\Tb)=\sqrt{\frac{k_{\Tb}}{\Tb}}\,\brk{\avg{x}_T-\avg{x}_{\Tb}}, \label{eq:ou1d-coefficients} \\
	\operatorname{He}_2(y)=y^2-1\quad&\longrightarrow\quad
	a_2(T,\Tb)=\frac{1}{\sqrt2}\Brk{\frac{k_{\Tb}}{\Tb}\brk{\frac{T}{k \brk{T}}-\frac{\Tb}{k_{\Tb}}}+a_1^2(T,\Tb)}. \label{eq:ou1d-coefficients_2}
\end{align}
The slowest coefficient $a_1$ is the displacement of the mean alone, while $a_2$ is the displacement of the second moment, which contains the variance and, through $a_1^2$, the mean as well.

\Cref{fig:bell}(b) shows both coefficients for our choice of mean and stiffness at $\Tb=1$, where $k_{\Tb}=1$ and $a_1$ is simply $\avg{x}_T-\avg{x}_{\Tb}$.
Because the equilibrium mean is nonmonotonic, so is $a_1$: it is largest at $\Tc=2$, the maximum of the mean, and besides $\Tb=1$ it vanishes again at $\Th\approx3.5$, where the mean returns to its bath value.
$\Th$ is thus a strong point, $T_s=\Th$, of a strong direct Mpemba effect, and any two preparations between $\Tc$ and $\Th$ also show a weak effect.
\Cref{fig:bell}(c) shows the resulting relaxation from $\Tc$ and $\Th$.
The hot preparation starts farther away ($D_0\approx0.2$ versus $0.12$), yet the two KL divergences, \cref{eq:kl}, cross at time $t_\times\approx0.23$, and at late times the hot one decays as $\e^{-4t}$ instead of $\e^{-2t}$, twice the rates $\lambda_2=2$ and $\lambda_1=1$.

The existence of the Mpemba effect is thus controlled by \emph{an equilibrium property}, the nonmonotonic temperature dependence of the mean, and can be read off from the equilibrium curve $\avg{x}_T$ without following the relaxation in time.
This link between equilibrium properties and the existence of the effect is the key insight of this work, and it carries over to the following sections.
Finally, if both the mean and the stiffness were independent of temperature, $a_1$ would vanish for every preparation, so the slowest relaxation would be controlled by $a_2$, the displacement of the variance $T/k$.
Since it grows monotonically with $T$, there is no Mpemba effect, as for the fixed spring of \cite{Walker2021}.

\section{Two dimensions: competing slow coordinates}
\label{sec:two-dimensional}

\begin{figure*}[t]
	\centering
	\includegraphics[width=\textwidth]{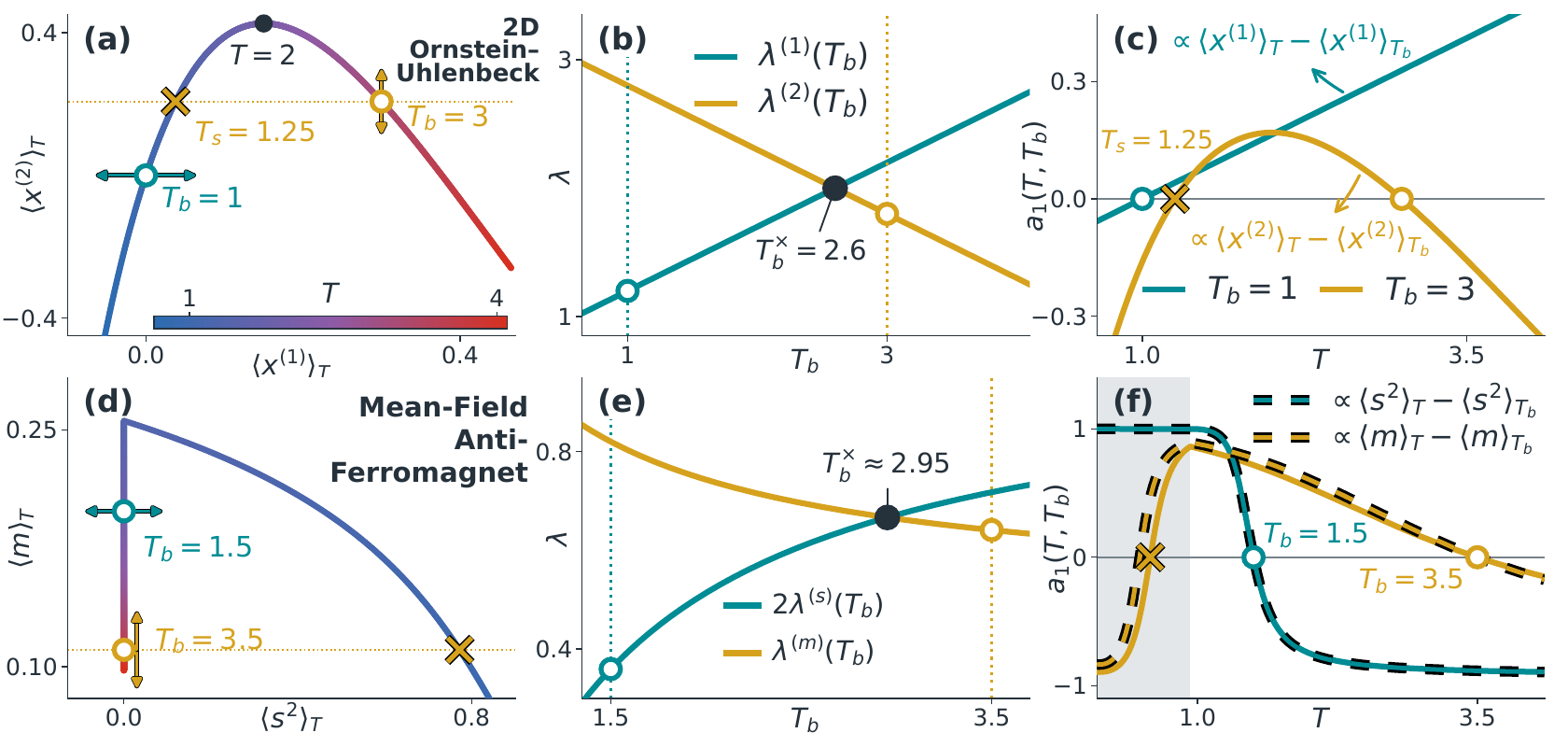}
	\caption{Rate crossing and the Mpemba effect in two dimensions: the solvable OU model, \cref{eq:uncoupled2d} (top), and the mean-field antiferromagnet at $H=0.5$ (bottom).
		Teal (gold) marks a bath below (above) the rate crossing $\Tb^{\times}$ (black dot); open circles mark the baths and crosses the strong points.
		(a) Equilibrium means along the preparations, colored by $T$; arrows mark the slow coordinate, and the black dot the maximum of $\avg{x^{(2)}}_T$.
		(b) Bath rates, \cref{eq:ou2d-coefficients}.
		(c) Slow coefficient, \cref{eq:ou2d-slow}, up to a positive factor: below $\Tb^{\times}$ it follows the monotonic $\avg{x^{(1)}}_T$, above it the nonmonotonic $\avg{x^{(2)}}_T$.
		(d) $\avg{m}_T$ against $\avg{s^2}_T$ for $N\to\infty$.
		(e) Rates, \cref{eq:afm-rates}, exact and equal to the harmonic ones, \cref{eq:stiffness}.
		(f) Slow coefficient rescaled to $[-1,1]$, exact for $N\to\infty$ (solid), \cref{eq:a1-limit}, and harmonic (dashed), \cref{eq:afmamplitudes}; gray marks antiferromagnetic preparations.}
	\label{fig:two-d}
\end{figure*}

In one dimension the story is simple: there is a single slow mode, its coefficient is the displacement of the mean, and a nonmonotonic mean is enough for a Mpemba effect.
Adding a second coordinate changes one thing: the two coordinates now compete for the slowest rate.
Each coordinate relaxes at its own rate, and since both rates depend on the bath temperature, the slowest rate can change as $\Tb$ is varied.
A nonmonotonic mean is therefore no longer sufficient for a Mpemba effect; it must also belong to the coordinate that relaxes slowest.
Below we follow how the effect appears and disappears as the slow coordinate switches.

To isolate this competition we choose two uncoupled coordinates,
\begin{equation}
\begin{aligned}
	 U_T(\bm x)&=\sum_{j=1,2}\frac{k^{(j)}(T)}{2}\,
	 \Brk{x^{(j)}-\avg{x^{(j)}}_T}^2,\\
	 \partial_tp\brk{x^{(1)},x^{(2)},t}&=\sum_{j=1,2}\partial_{x^{(j)}}\!\Brk{k^{(j)}(\Tb)\brk{x^{(j)}-\avg{x^{(j)}}_{\Tb}}p+\Tb\,\partial_{x^{(j)}}p},
\end{aligned}
	 \label{eq:ou2d-potential}
\end{equation}
with means and stiffnesses
\begin{equation}
	\begin{aligned}
		 \avg{x^{(1)}}_T&=0.15\,\brk{T-1}, &
		 k^{(1)}(T)&=0.7+0.5\,T,\\
		 \avg{x^{(2)}}_T&=2\Brk{T\e^{-0.5\brk{T-1}}-1}, &
		 k^{(2)}(T)&=3.3-0.5\,T,
	\end{aligned}
	 \label{eq:uncoupled2d}
\end{equation}
where $0<T<6.6$, so that both stiffnesses are positive.  The first mean is monotonic.  The second
grows up to $T=2$ and then decreases, so the preparation curve in
\cref{fig:two-d}(a) is nonmonotonic only along $x^{(2)}$.  As in one
dimension, the initial KL divergence ($D_0$ in \cref{eq:kl}) grows monotonically with $\abs{T-\Tb}$
(see \cref{app:kl-monotonic}).

Since the coordinates are uncoupled, each relaxes independently according to \cref{eq:ou1d-fp}, with its own mean and stiffness.
Applying \cref{eq:ou1d-coefficients} to each coordinate gives the two candidates for the slowest mode, with coefficients and rates
\begin{equation}\label{eq:ou2d-coefficients}
	a^{(j)}(T,\Tb)=\sqrt{\frac{k^{(j)}(\Tb)}{\Tb}}\,\brk{\avg{x^{(j)}}_T-\avg{x^{(j)}}_{\Tb}}, \qquad
	\lambda^{(j)}(\Tb)=k^{(j)}(\Tb). 
\end{equation}
The slow coefficient is the one with the smaller rate,
\begin{equation}
	a_1(T,\Tb)=
	\begin{cases}
		a^{(1)}(T,\Tb), & \lambda^{(1)}(\Tb)<\lambda^{(2)}(\Tb),\\[2pt]
		a^{(2)}(T,\Tb), & \lambda^{(1)}(\Tb)>\lambda^{(2)}(\Tb).
	\end{cases}
	\label{eq:ou2d-slow}
\end{equation}
The preparation temperature $T$ thus affects $a_1$ only through the equilibrium means, while the bath temperature $\Tb$ selects which of the two means sets $a_1$.

\Cref{fig:two-d}(b) shows that the two rates cross at $\Tb^{\times}=2.6$, and this crossing is where the behavior of the slow coefficient $a_1$ changes.
Below the crossing, for $\Tb < \Tb^{\times}$, $x^{(1)}$ is the slow coordinate and controls $a_1$.
Since $\avg{x^{(1)}}_T$ is monotonic in temperature, so is $a_1$, and there is no Mpemba effect.
This is exemplified for $\Tb=1$ by an open circle in \cref{fig:two-d}(b), where $\lambda^{(1)}<\lambda^{(2)}$, and by the corresponding $a_1$ in \cref{fig:two-d}(c).
Above the crossing, for $\Tb > \Tb^{\times}$, $x^{(2)}$ takes over as the slow coordinate and controls $a_1$, and the nonmonotonicity of $\avg{x^{(2)}}_T$ now induces a Mpemba effect.
This is exemplified for $\Tb=3$ by an open circle in \cref{fig:two-d}(b), where $\lambda^{(2)}<\lambda^{(1)}$, and by the corresponding $a_1$ in \cref{fig:two-d}(c).
There, besides $\Tb$, $a_1$ vanishes at $T_s\approx1.25$, where $\avg{x^{(2)}}_T$ returns to its bath value (crosses in \cref{fig:two-d}(a),(c)), so the preparation at $T_s$ relaxes at the rate $\lambda^{(1)}$ instead of $\lambda^{(2)}$: a strong inverse Mpemba effect.
Preparations between $T_s$ and the maximum of $\avg{x^{(2)}}_T$ at $T=2$ also show a weak effect.
This establishes the connection between equilibrium properties and the existence of the Mpemba effect in two dimensions as well, with the rate crossing marking the onset of the effect.
Coupling the coordinates would only rotate the normal directions and turn the crossing into an avoided one, and the same strong effect survives (see \cref{sec:general} and \cref{app:avoided}).

\section{Mean-field Ising antiferromagnet}
\label{sec:antiferromagnet}

We now show that the same picture captures the essence of the Mpemba effect in a many-body system, the mean-field Ising antiferromagnet.
It is a minimal many-body model: being mean-field, it is fully described for $N\to\infty$ by a free energy of two macroscopic order parameters, much as the OU models of \cref{sec:one-dimensional,sec:two-dimensional} are described by a potential.
Its relaxation, however, is nonlinear and is not an OU process, and it shows both direct and inverse Mpemba effects, depending on $\Tb$ \cite{Klich2019}.
The question is whether the rule of thumb survives: whether the temperature dependence of an equilibrium order parameter still decides where the effect appears.

The system consists of two sublattices of $N/2$ Ising spins each, with an antiferromagnetic coupling between every pair of spins on different sublattices and a field $H\geq0$ \cite{Vives1997}.
The coupling favors opposite magnetizations on the two sublattices, while the field favors spins along it.
Since every spin interacts equally with all the spins of the other sublattice, the state is described by two macroscopic variables.
A macrostate is fixed by the magnetization per spin, $m$, and the staggered magnetization, $s$, and has energy $E$ and entropy $S$,
\begin{equation}
	\begin{aligned}
		m&=\frac{\brk{N^{(1)}_\uparrow-N^{(1)}_\downarrow}+\brk{N^{(2)}_\uparrow-N^{(2)}_\downarrow}}{N}, &
		s&=\frac{\brk{N^{(1)}_\uparrow-N^{(1)}_\downarrow}-\brk{N^{(2)}_\uparrow-N^{(2)}_\downarrow}}{N},\\
		E(m,s)&=\frac N2\brk{m^2-s^2}-NHm, &
		S(m,s)&=N\ln2-\frac N4\Brk{f(m+s)+f(m-s)},
	\end{aligned}
	\label{eq:ms}
\end{equation}
where $N^{(k)}_{\uparrow,\downarrow}$ count the up and down spins on sublattice $k=1,2$, and $f(x)=\brk{1+x}\ln(1+x)+\brk{1-x}\ln(1-x)$.
Both $m$ and $s$ lie between $-1$ and $1$.
The magnetization $m$ is the average spin, and $m=1$ ($-1$) when all spins point up (down).
The staggered magnetization $s$ is half the difference between the magnetizations of the two sublattices: it vanishes when they are equally magnetized, and $s=\pm1$ when each is fully polarized, in opposite directions.
The free energy $\mathcal F(m,s,T)=E(m,s)-T\,S(m,s)$ gives the equilibrium distribution over macrostates, $\pi(m,s;T)\propto\e^{-\mathcal F(m,s,T)/T}$.
For low enough temperatures, this system exhibits a phase transition from a paramagnetic phase (with vanishing $s$) to an antiferromagnet (with nonzero $s$) \cite{Vives1997}.

Following \cite{Klich2019}, we evolve the system with single-spin-flip Glauber dynamics, in which a spin on sublattice $i$ flips at a rate set by the field of the other sublattice, $j$ (rates in \cref{app:afm-ninf}).
To leading order in $1/N$, the master equation of these rates becomes the Fokker--Planck equation for $p \brk{m,s,t}$
\begin{equation}
	\begin{aligned}
	 \partial_tp=&\frac12\,\partial_m\Brk{\brk{m-\frac{\tau\bigbrk{m-s}+\tau\bigbrk{m+s}}{2}}p} +\frac12\,\partial_s\Brk{\brk{s-\frac{\tau\bigbrk{m-s}-\tau\bigbrk{m+s}}{2}}p}\\
	 &+\frac1{4N}\sum_{\sigma =\pm1}\brk{\partial_m+\sigma\partial_s}^2\Brk{p-\brk{m+\sigma s}\tau\bigbrk{m-\sigma s}p},
	\end{aligned}
	 \label{eq:glauber-fp}
\end{equation}
where $\tau\bigbrk{x}=\tanh\Brk{\brk{H-x}/\Tb}$.
It takes the place of \cref{eq:master}, and its stationary state is the bath equilibrium, $\pi(m,s;\Tb)$.

\subsection{Slow mode and the rule of thumb}
\label{sec:afm-exact}\label{sec:afm-ou}

We first approximate the relaxation near the bath by an OU process.
Expanding \cref{eq:glauber-fp} to second order around a paramagnetic bath, $m=\avg{m}_{\Tb}$ and $s=0$, that is, linearizing the drift and freezing the diffusion at its bath value, gives
\begin{equation}
 \partial_tp={}\lambda^{(m)}\partial_m\Brk{\brk{m-\avg{m}_{\Tb}}p}+\lambda^{(s)}\partial_s\brk{s p}
 +\frac{1-\avg{m}_{\Tb}^2}{2N}\brk{\partial_m^2+\partial_s^2}p,
 \label{eq:ou-fp}
\end{equation}
with the rates
\begin{equation}
	\lambda^{(m)}(\Tb)=\frac12\Brk{1+\frac{1-\avg{m}_{\Tb}^2}{\Tb}},\qquad
	\lambda^{(s)}(\Tb)=\frac12\Brk{1-\frac{1-\avg{m}_{\Tb}^2}{\Tb}}.
	\label{eq:afm-rates}
\end{equation}
We note that the rates are set by the free-energy curvatures at the bath,
\begin{equation}
	 \lambda^{(j)}=\frac{1-\avg{m}_{\Tb}^2}{2N\Tb}\,\partial_j^2\mathcal F,\qquad j=m,s,
	 \label{eq:stiffness}
\end{equation}
so the OU process is fixed by the free-energy curvatures and the mobility at the bath; since the mobility is the same for $m$ and $s$, which of them relaxes more slowly is decided by the curvatures alone (\cref{sec:general}).

This is the uncoupled OU process of \cref{sec:two-dimensional}, with $m$ and $s$ in place of $x^{(1)}$ and $x^{(2)}$, so the intuition of that section carries over.
The one new ingredient is symmetry: exchanging the two sublattices, $s\mapsto-s$, leaves $\mathcal F$ and the dynamics unchanged.
Each preparation is an equilibrium state \cite{Klich2019}, and is therefore symmetric too, so $\avg{s}$ vanishes initially and stays zero throughout the relaxation.
The leading moment of $s$ is then $\avg{s^2}$, which, like the second moment in \cref{sec:one-dimensional}, relaxes at $2\lambda^{(s)}$, twice the rate of $s$ (the slow mode when the symmetry is broken \cite{Blom2026}; see \cref{app:afm-ninf}).
The two moments that play the role of the two means of \cref{sec:two-dimensional} are therefore $\avg{m}$, which relaxes at $\lambda^{(m)}$, and $\avg{s^2}$, which relaxes at $2\lambda^{(s)}$.

Since each moment of an OU process relaxes exponentially, the slow coefficient is the displacement of the slower of the two, as in \cref{eq:ou2d-slow},
\begin{equation}
	a_1(T,\Tb)=
	\begin{cases}
		a^{(m)}(T,\Tb)\propto\avg{m}_T-\avg{m}_{\Tb}, & \lambda^{(m)}(\Tb)<2\lambda^{(s)}(\Tb),\\[4pt]
		a^{(s^2)}(T,\Tb)\propto\avg{s^2}_T-\avg{s^2}_{\Tb}, & \lambda^{(m)}(\Tb)>2\lambda^{(s)}(\Tb),
	\end{cases}
	\label{eq:afmamplitudes}
\end{equation}
up to positive factors.
An effect therefore requires the slow moment to be \emph{nonmonotonic} in the preparation temperature $T$, and which of the two moments is nonmonotonic depends on the field.
For $H<1$, $\avg{s^2}_T$ decreases monotonically with $T$, while $\avg{m}_T$ first grows and then decreases, turning over at the transition (\cref{fig:two-d}(d), at $H=0.5$; equilibrium values at the global minimum of $\mathcal F(m,s,T)$ for $N\to\infty$, \cref{app:afm-details}).
At these fields the rule of thumb therefore predicts no effect at baths where $s^2$ is slow, and an effect where $m$ is.
At $H=0.5$, $\lambda^{(m)}$ and $2\lambda^{(s)}$ cross at $\Tb^{\times}\approx2.95$ (\cref{fig:two-d}(e)), so $s^2$ is slow below the crossing and $m$ above it.
Accordingly, at $\Tb=1.5<\Tb^{\times}$ the slow coefficient $a_1$ is monotonic and there is no effect, while at $\Tb=3.5>\Tb^{\times}$ it vanishes at the cold preparation $T_s\approx0.48$, where $\avg{m}_T$ returns to its bath value: a strong inverse Mpemba effect (dashed in \cref{fig:two-d}(f)).
For $H>1$ the roles are reversed: $\avg{s^2}_T$ becomes the nonmonotonic moment, and the effect appears at baths where $s^2$ is slow (\cref{sec:phase-map}).

The OU process is only an approximation: the drift of \cref{eq:glauber-fp} is nonlinear, and the preparations may lie far from the bath.
To test it, we find the slow mode exactly for $N\to\infty$.
By \cref{eq:expansion}, any observable $O$ relaxes as a sum over the modes, $\avg{O}_t-\avg{O}_{\Tb}=\sum_n a_n\e^{-\lambda_n t}O_n$, with $O_n=\int\dd m\,\dd s\,O\,v_n$.
At late times only the slowest modes survive, and $\avg{O}_t$ is close to its bath value, so the rates of these modes follow from the equations of motion of the moments linearized around the bath.
Since $\avg{s}$ vanishes by symmetry, the slow mode appears either in $\avg{m}$ or in $\avg{s^2}$, so we need to linearize only their equations, and this gives back exactly the rates of \cref{eq:afm-rates}: $\avg{m}$ relaxes at $\lambda^{(m)}$ and $\avg{s^2}$ at $2\lambda^{(s)}$ (full details in \cref{app:afm-ninf}).
This is no coincidence: the OU process is itself the linearization of the dynamics around the bath, and late times probe only this neighborhood.
The slow moment for $N\to\infty$ is therefore selected exactly as in the OU process.

With the rates exact, only the slow coefficient can differ from its harmonic value.
Taking $O=m$ or $O=s^2$ in the mode sum above, we find that only the slow mode survives at late times $\avg{O}_t-\avg{O}_{\Tb}\simeq a_1\e^{-\lambda_1 t}O_1$, so the slow coefficient is the amplitude of the slow moment that remains at late times,
\begin{equation}
	a_1(T,\Tb) = 
	\begin{cases}
		a^{(m)} \propto \lim\limits_{t\to\infty}\e^{\lambda^{(m)}t}\Brk{\avg{m}_t-\avg{m}_{\Tb}}, & \lambda^{(m)}(\Tb)<2\lambda^{(s)}(\Tb),\\[4pt]
		a^{(s^2)} \propto \lim\limits_{t\to\infty}\e^{2\lambda^{(s)}t}\Brk{\avg{s^2}_t-\avg{s^2}_{\Tb}}, & \lambda^{(m)}(\Tb)>2\lambda^{(s)}(\Tb),
	\end{cases}
	\label{eq:a1-limit}
\end{equation}
up to factors independent of $T$, where $\avg{m}_t$ and $\avg{s^2}_t$ evolve from the preparation at $T$.
In the OU process each moment relaxes exponentially at all times, at its own rate, $\avg{O}_t-\avg{O}_{\Tb}=\e^{-\lambda^{(O)}t}\brk{\avg{O}_T-\avg{O}_{\Tb}}$, and \cref{eq:a1-limit} reduces to the displacements of \cref{eq:afmamplitudes}.
In the exact dynamics the moments relax exponentially only near the bath, so the coefficient also depends on the nonlinear path from the preparation to the bath.

\Cref{fig:two-d}(f) compares the two coefficients at $H=0.5$.
At $\Tb=1.5$, where $s^2$ is slow but the bath is far from the crossing, they are almost indistinguishable.
At $\Tb=3.5$ both coefficients vanish at a cold preparation ($T_s\approx0.48$ harmonic, $0.58$ exact), so the harmonic approximation correctly predicts the strong inverse effect, even though the strong preparation lies far from the bath, in the antiferromagnetic phase (gray in \cref{fig:two-d}(f)).
As in the solvable model of \cref{sec:two-dimensional}, the effect thus follows from the equilibrium moments alone: $\avg{s^2}_T$ is monotonic and $\avg{m}_T$ is not, with $s^2$ and $m$ playing the roles of $x^{(1)}$ and $x^{(2)}$, respectively.

\begin{figure*}[t]
	\centering
	\includegraphics[width=\textwidth]{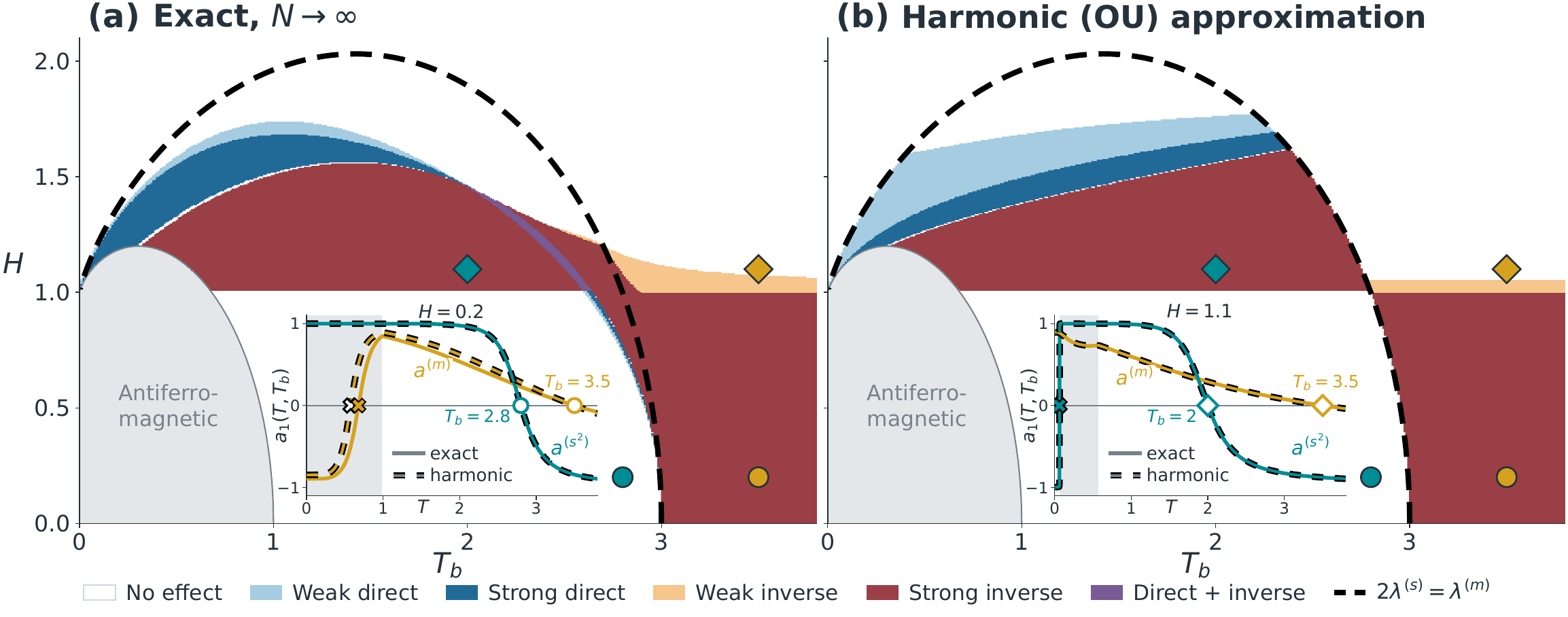}
	\caption{Mpemba phase diagram of the mean-field antiferromagnet for $N\to\infty$: (a) exact, \cref{eq:a1-limit}, and (b) harmonic, \cref{eq:afmamplitudes}, on the same grid of baths $(\Tb,H)$, each classified by how its slow coefficient depends on the preparation temperature (\cref{sec:phase-map}).
		Dashed: the rate crossing $2\lambda^{(s)}=\lambda^{(m)}$, identical in both panels; gray: the antiferromagnetic phase, not classified.
		Filled symbols, the same in both panels, mark four baths: circles at $H=0.2$ ($\Tb=2.8$ and $3.5$) and diamonds at $H=1.1$ ($\Tb=2$ and $3.5$), colored by the slow moment ($s^2$ teal, $m$ gold).
		Insets: $a_1(T,\Tb)$ at the circle baths ($H=0.2$) and at the diamond baths ($H=1.1$), each comparing the exact (solid) and harmonic (dashed) coefficients, rescaled as in \cref{fig:two-d}(f) and labeled by the slow coefficient, $a^{(m)}$ or $a^{(s^2)}$; open symbols mark the bath, crosses the strong points, and gray bands antiferromagnetic preparations.}
	\label{fig:phase-diagram}
\end{figure*}

\subsection{Reconstructing the Mpemba phase diagram}
\label{sec:phase-map}

The two baths of \cref{fig:two-d} are points of a full phase diagram in the $(\Tb,H)$ plane, which maps where the effect is absent, weak or strong, and whether it is direct or inverse.
We use this diagram to test the rule of thumb, and show that the harmonic approximation, which uses equilibrium properties alone, reproduces it well.
\Cref{fig:phase-diagram} compares the diagram obtained from the exact coefficients for $N\to\infty$, from \cref{eq:a1-limit} (panel (a)), with the one from the harmonic approximation, \cref{eq:afmamplitudes} (panel (b)).
Each bath $(\Tb,H)$ is classified by the dependence of its slow coefficient $a_1(T,\Tb)$ on the preparation temperature $T$: the effect is strong if $a_1$ vanishes at some $T\neq\Tb$, weak if $\abs{a_1}$ is nonmonotonic in $T$, and direct or inverse for $T>\Tb$ or $T<\Tb$, respectively.
The only baths we leave out are those in the antiferromagnetic phase, where the free energy has two symmetry-broken minima and the slow mode requires a separate analysis.

At every field, the diagram follows the picture of \cref{sec:afm-exact}: the rate crossing (dashed), identical in both panels since the harmonic rates are exact, decides which moment is slow, and the monotonicity of that moment in $T$ decides whether there is an effect.
For $H<1$, where $\avg{s^2}_T$ is monotonic and $\avg{m}_T$ is not, baths inside the dashed curve, where $s^2$ is slow, show no effect, while baths outside it, where $m$ is slow, show a strong inverse effect, since a cold preparation shares the value of $\avg m$ with the bath (the inset of \cref{fig:phase-diagram}(a) compares the harmonic and exact slow coefficients at two baths with $H=0.2$).
The exact diagram (\cref{fig:phase-diagram}(a)) is nearly identical in this range.
The two coincide exactly at the onset of the strong inverse region, $\Tb=3$ as $H\to0$, which is fixed by the crossing alone, and they deviate only in a narrow strip along the crossing.
Elsewhere the agreement is excellent, in particular over the strong inverse region at high $\Tb$, where the nonlinear path shifts the temperature of the strong point but not its existence.

For $H>1$ the roles are reversed.
The field now fully polarizes the ground state, so $\avg{s^2}_T$ vanishes at low $T$, peaks at intermediate $T$ and decreases at high $T$, while $\avg{m}_T$ becomes monotonic, apart from a narrow window near $H=1$ (see more details in \cref{app:disagreement}).
The effect therefore appears inside the dashed curve, where $s^2$ is slow, and not outside it.
The inset of \cref{fig:phase-diagram}(b) compares the harmonic and exact slow coefficients at two baths with $H=1.1$.
At $\Tb=2$ the slow coefficient is $a^{(s^2)}$, and it vanishes at a cold preparation: a strong inverse effect.
At $\Tb=3.5$ it is $a^{(m)}$, which is monotonic, so there is no effect.
The exact diagram has the same structure, but differs more from the harmonic one than for $H<1$ (see more details in \cref{app:disagreement}).

The diagram highlights two points.
First, the controlling moment changes with $H$, and which one it is follows from the rates rather than being chosen in advance.
For $H>1$ it is $N\avg{s^2}_T=T\chi_s$, the staggered susceptibility used as a proxy for the slow mode in \cite{GonzalezAdalid2024,Pomares2026}.
For $H<1$, however, it is the magnetization, which is not a susceptibility, so the effect need not come from a nonmonotonic susceptibility.
Second, the effect does not require a phase transition: above $H\approx1.2$ the antiferromagnet is paramagnetic at every temperature, yet both diagrams show strong direct and inverse effects up to $H\approx1.7$, because $\avg{s^2}_T$ remains nonmonotonic there.
Unlike in \cite{Blom2026}, where the strong effect comes from the staggered order and arises through re-entrance, here it comes from the staggered fluctuations and requires no ordered phase at all.
Together, these show that the effect is tied neither to a particular observable nor to critical behavior: what decides it is whether the moment selected by the bath is nonmonotonic in the preparation temperature, whatever the origin of that nonmonotonicity is.

\section{The general case: coupled coordinates in any dimension}
\label{sec:general}\label{sec:ou}\label{sec:model-a}

The antiferromagnet was simple in one respect: symmetry made $m$ and $s$ the normal coordinates of the bath, so its harmonic approximation was the uncoupled OU process of \cref{sec:two-dimensional}.
The same picture can arise in any mean-field system, in which the coordinates are generally coupled.
Near the bath its relaxation is approximated by an OU process (\cref{fig:mechanism}(b)), which is not exact far from the bath, but serves as a rule of thumb: it identifies the equilibrium moment that controls the slow relaxation, and its temperature dependence indicates whether a Mpemba effect occurs.

Let $\bm x$ collect the $d$ intensive macroscopic variables of the system, such as $m$ and $s$, with $d$ fixed as $N$ grows, and let $\mathcal F(\bm x,T)$ be their free energy.
We take the relaxation to be of the common Model-A form \cite{Hohenberg1977}, a noisy gradient flow on the free energy of the bath,
\begin{equation}
	\partial_tp(\bm x,t)=\nabla\!\cdot\!\Bigl\{\bm\mu(\bm x)\Brk{p\,\nabla\mathcal F(\bm x,\Tb)
		+\Tb\nabla p}\Bigr\},
	\label{eq:mf-fp}
\end{equation}
which is \cref{eq:master} with the potential replaced by the free energy, and with a symmetric, positive mobility matrix $\bm\mu(\bm x)=O(N^{-1})$.
For the antiferromagnet with Glauber dynamics it reproduces the relaxation near the bath, \cref{eq:ou-fp}.
Expanding to second order around the bath minimum, as in \cref{sec:afm-ou}, gives the OU process
\begin{equation}
 \partial_tp(\bm x,t)=\nabla\!\cdot\!\Bigl\{\bm\mu_b\Brk{\bm K_b\brk{\bm x-\avg{\bm x}_{\Tb}}p
 +\Tb\nabla p}\Bigr\},
 \label{eq:ou-fp-general}
\end{equation}
where $\bm K_b$ is the free-energy Hessian at the minimum, with elements $\brk{\bm K_b}_{ij}=\partial_{x^{(i)}}\partial_{x^{(j)}}\mathcal F(\bm x,\Tb)\big|_{\bm x=\avg{\bm x}_{\Tb}}$ (\cref{fig:mechanism}(b)), and $\bm\mu_b=\bm\mu(\avg{\bm x}_{\Tb})$ is the mobility there.
In the antiferromagnet both matrices were diagonal, \cref{eq:stiffness}; in general they are not, and the coordinates are coupled.

Coupling, however, only rotates the coordinates.
The eigenvectors of $\bm\mu_b\bm K_b$ define $d$ normal coordinates $y^{(j)}$, $j=1,\ldots,d$: linear combinations of the components of $\bm x-\avg{\bm x}_{\Tb}$, each scaled to unit variance in the bath, like the standardized coordinate $y$ of \cref{sec:one-dimensional}.
Each relaxes independently, with a rate $\lambda^{(j)}(\Tb)$ given by the corresponding eigenvalue of $\bm\mu_b\bm K_b$.
Each mode is then a product of the Hermite polynomials of \cref{eq:ou1d-spectrum}, one for each normal coordinate, and is labeled by their degrees $n_j=0,1,2,\ldots$, so that
\begin{equation}
 \lambda_{n_1\ldots n_d}=\sum_{j=1}^{d}n_j\lambda^{(j)},\qquad
 a_{n_1\ldots n_d}=\avg{\prod_{j=1}^{d}\frac{\operatorname{He}_{n_j}\bigbrk{y^{(j)}}}{\sqrt{n_j!}}}_T ,
 \label{eq:ou-spectrum}
\end{equation}
and, as in one dimension, every coefficient is a combination of \emph{equilibrium moments} at the preparation temperature.
The lowest ones are a first, a second, and a mixed moment,
\begin{equation}
\begin{aligned}
 a^{(j)}&=\avg{y^{(j)}}_T, && \lambda^{(j)},\\
 a^{(j^2)}&=\Brk{\avg{\bigbrk{y^{(j)}}^2}_T-1}/\sqrt2, && 2\lambda^{(j)},\\
 a^{(ij)}&=\avg{y^{(i)}y^{(j)}}_T, && \lambda^{(i)}+\lambda^{(j)}.
\end{aligned}
 \label{eq:lowest}
\end{equation}
For the uncoupled examples the first line reduces to \cref{eq:ou1d-coefficients,eq:ou2d-coefficients}, and for the antiferromagnet, with $y^{(m)}\propto m-\avg{m}_{\Tb}$ and $y^{(s)}\propto s$, $a^{(m)}$ and $a^{(s^2)}$ are the two candidates of \cref{eq:afmamplitudes}; for symmetry-broken preparations, $a^{(s)}$ is the amplitude found in the pair approximation in \cite{Blom2026}.

The rule is therefore the same in any dimension: the bath selects which moment is slow, and the preparation enters only through its \emph{equilibrium value}.
The slow coefficient $a_1$ is the lowest-rate coefficient in \cref{eq:ou-spectrum} that does not vanish identically.
Usually this is the first moment with the smallest $\lambda^{(j)}$, as in \cref{sec:two-dimensional}.
Symmetry can remove it, as in the antiferromagnet: if $\avg{y^{(j)}}_T=0$ for every $T$, the second moment $a^{(j^2)}$ competes with the first moments of the other coordinates.
Mixed moments never compete, since each decays faster than the second moment of its softer coordinate, $\lambda^{(i)}+\lambda^{(j)}>2\min\brk{\lambda^{(i)},\lambda^{(j)}}$.
Once $a_1$ is identified, the existence of the effect is again an \emph{equilibrium} question: the selected moment must be nonmonotonic in $T$, and a second zero of it gives a strong effect.
Nothing in this selection refers to an order parameter, a susceptibility or a phase transition: the controlling moment follows from $\bm\mu_b\bm K_b$ and the symmetry of the preparations alone.
The only new feature of coupling is that the normal coordinates keep rotating as $\Tb$ is varied, so a crossing of two rates generically becomes an avoided one.
The rule then applies to the moments of the slow normal coordinate on either side of it, and a coupled two-dimensional example (see \cref{app:avoided}) shows the same strong effect as its uncoupled counterpart.

\section{Discussion and outlook}

We have proposed a simple rule of thumb for anticipating the Markovian Mpemba effect: find the moment that relaxes slowest at the bath, and check whether its equilibrium value is nonmonotonic in the preparation temperature.
If it is, an effect is expected, and it is strong where the moment returns to its bath value.
The rule separates the roles of the bath and of the preparation.
The bath, through the free-energy curvature and the mobility, decides which moment relaxes slowest, while the preparation enters only through the equilibrium value of that moment.
A switch of the slow moment or of the relaxation direction \cite{Walker2021,Teza2023} therefore does not by itself create or remove the effect; only the monotonicity of the newly selected moment matters.
The rule requires only equilibrium properties, and only a few macroscopic moments rather than the full distribution, without solving the dynamics or diagonalizing evolution operators whose size grows rapidly with the system; for the same reason, it can be tested directly in macroscopic experiments.
In the OU process this picture is exact and complete: the coefficient of the slowest mode is the displacement of an equilibrium mean from its bath value, and its temperature dependence alone decides whether there is an effect, whether it is direct or inverse, and where its strong points lie.

In mean-field systems the rule becomes an approximation, obtained by expanding the free energy around the bath minimum into an OU process.
Although this expansion is controlled only near the bath, while the preparations may lie far from it, we showed that it makes for an excellent approximation for the mean-field Ising antiferromagnet: the harmonic rates are exact, and the harmonic phase diagram captures the main structure of the exact one, and is almost indistinguishable from it for $H<1$.
The antiferromagnet thus shows that neither a phase transition nor a susceptibility is required: the rule derives the controlling moment itself. 
Where $s^2$ is slow, it recovers the susceptibility proxy of \cite{GonzalezAdalid2024,Pomares2026}; where $m$ is slow, it selects the uniform magnetization instead. 
It also predicts strong effects where there is no transition at all.
Finally, we showed that the same rule applies to any mean-field system with Model-A dynamics: the free-energy Hessian and the mobility at the bath give the normal coordinates and their rates, and the slowest of their low-order moments controls the effect.
Natural next steps are to extend the rule to baths in ordered phases, where the free energy has several minima, and to conserved, inertial \cite{AharonyShapira2026} or quantum \cite{Ares2025review} dynamics.
Since the rule locates the strong preparations from equilibrium data alone, it may also guide the design of protocols for faster relaxation \cite{GalRaz2020,Boundary2023, GonzalezAdalid2024}.

\begin{acknowledgments}
	We thank Oren Raz and Gianluca Teza for many discussions and for their guidance, which shaped this work.
	We also thank Bara Levit, Adam Hitin Bialus and Harel Kol-Namer for reading the manuscript and for helpful comments and discussions.
\end{acknowledgments}

\appendix

\renewcommand{\thefigure}{\thesection\arabic{figure}}
\renewcommand{\theHfigure}{\thesection\arabic{figure}}

\section{Monotonic initial distance}
\label{app:kl-monotonic}
\setcounter{figure}{0}

The Mpemba effect is stated in terms of temperatures, but what makes it anomalous is a comparison of distances: the preparation that relaxes faster must start farther from equilibrium.
The temperature is a faithful proxy for this distance only if the initial divergence $D_0(T)$ grows monotonically with $\abs{T-\Tb}$.
If it did not, a hotter preparation could simply start closer to the bath, and its faster relaxation, even a vanishing slow coefficient, would be no more surprising than that of a preparation at $\Tb$ itself.

\begin{figure*}[t]
\centering
\includegraphics[width=\textwidth]{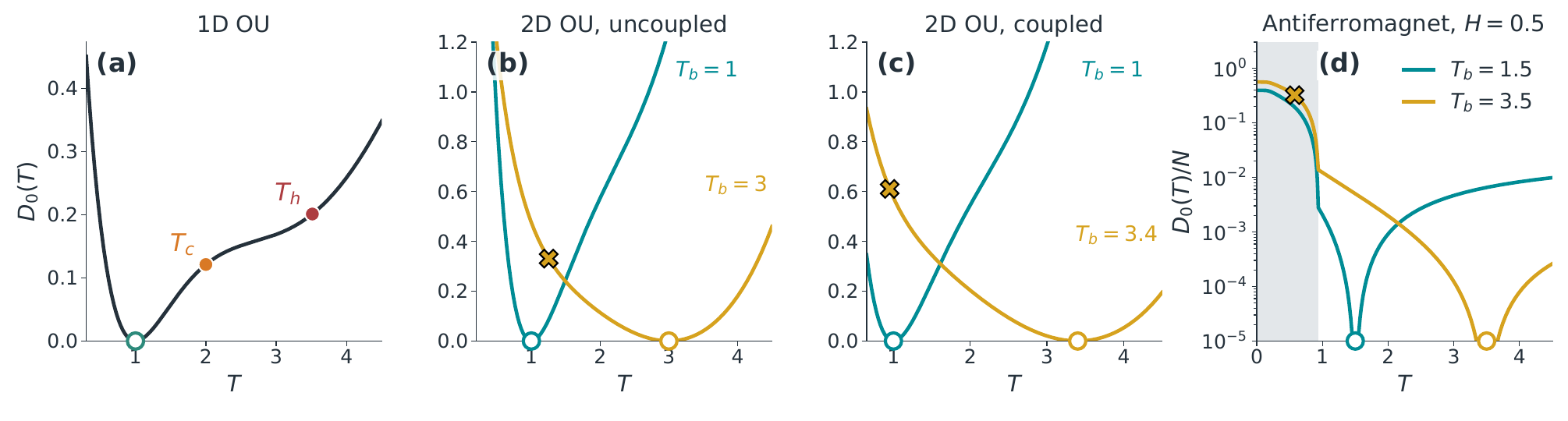}
\caption{Initial Kullback--Leibler divergence $D_0(T)$ between the preparation at $T$ and the bath, for every model of the main text; open circles mark the bath, and crosses the strong points.
(a) One-dimensional OU process, \cref{eq:bellmodel}, at $\Tb=1$, with the preparations $\Tc$ and $\Th$ of Fig.~\ref{fig:bell}.
(b) Uncoupled two-dimensional OU process, \cref{eq:uncoupled2d}, at the baths of Fig.~\ref{fig:two-d}(a)--(c).
(c) Coupled two-dimensional OU process, \cref{eq:avoided}, at the baths of Fig.~\ref{fig:avoided}.
(d) Antiferromagnet at $H=0.5$, per spin for $N\to\infty$, \cref{eq:kl-afm}, at the baths of Fig.~\ref{fig:two-d}(d)--(f) (logarithmic scale); the gray band marks antiferromagnetic preparations.
In every case $D_0$ grows monotonically as $T$ moves away from the bath.}
\label{fig:kl}
\end{figure*}

For a system whose energy $E(x)$ does not depend on temperature, monotonicity is guaranteed.
With $\pi(x;T)\propto\e^{-E(x)/T}$, a direct calculation gives
\begin{equation}
	\frac{\dd D_0}{\dd T}=\frac{T-\Tb}{T\,\Tb}\,C(T),
	\label{eq:kl-canonical}
\end{equation}
where $C(T)=\dd\avg{E}_T/\dd T=\mathrm{Var}_T(E)/T^2\geq0$ is the heat capacity at temperature $T$, so $D_0$ decreases for $T<\Tb$ and increases for $T>\Tb$.
This covers the antiferromagnet.
Its free energy $\mathcal F(m,s,T)$ does depend on $T$, but only through the entropy of the macrostates; the energy $E(m,s)$ of \cref{eq:ms} does not, and the preparations are canonical distributions of this energy over the microstates (see below).
The OU models, however, are defined by effective potentials $U_T(x)$ whose centers and stiffnesses depend on $T$, as coarse-grained free energies generally do.
For them \cref{eq:kl-canonical} does not hold, and a strong enough temperature dependence of the mean could make $D_0$ nonmonotonic.
We therefore check every model of the main text explicitly.
In all of them $D_0$ increases monotonically as the preparation temperature moves away from the bath (Fig.~\ref{fig:kl}), so their strong points are genuine Mpemba effects and not an artifact of the choice of distance.

\paragraph*{One dimension.}
The preparations of \cref{eq:bellmodel} are Gaussian,
\begin{equation}
	\pi(x;T)=\sqrt{\frac{k(T)}{2\pi T}}\,\exp\!\Brk{-\frac{k(T)}{2T}\bigbrk{x-\avg{x}_T}^2},
	\label{eq:pi-1d}
\end{equation}
with mean $\avg{x}_T=T\e^{1-T/2}$ and variance $\mathrm{Var}_T(x)=T/k(T)$, $k(T)=\brk{1+4T^2}/5$; both depend on $T$.
The divergence between two such Gaussians is
\begin{equation}
	D_0(T)=\frac12\Brk{\bigbrk{r-1-\ln r}+\frac{\bigbrk{\avg{x}_T-\avg{x}_{\Tb}}^2}{\mathrm{Var}_{\Tb}(x)}},\qquad r=\frac{\mathrm{Var}_T(x)}{\mathrm{Var}_{\Tb}(x)},
	\label{eq:initialKL1d}
\end{equation}
where $r$ is the ratio of the preparation and bath variances.
Both terms are manifestly nonnegative: the first, since $\ln r\leq r-1$ with equality only at $r=1$, measures the mismatch of the widths, and the second, the squared displacement of the mean in units of the bath variance.
At $\Tb=1$, where $\mathrm{Var}_{\Tb}(x)=1$, $D_0$ decreases strictly for $T<1$ and increases strictly for $T>1$ (Fig.~\ref{fig:kl}(a)).
In particular, $D_0(\Tc)\approx0.12$ and $D_0(\Th)\approx0.20$: the hotter preparation starts farther away, yet the two KL divergences cross at $t_\times\approx0.23$, where both are about $0.06$.

\paragraph*{Two dimensions.}
For \cref{eq:uncoupled2d} the coordinates are independent, so $D_0$ is the sum of \cref{eq:initialKL1d} over the two coordinates,
\begin{equation}
	D_0(T)=\frac12\sum_{j=1,2}\Brk{\bigbrk{r_j-1-\ln r_j}+\frac{\bigbrk{\avg{x^{(j)}}_T-\avg{x^{(j)}}_{\Tb}}^2}{\mathrm{Var}_{\Tb}(x^{(j)})}},\qquad r_j=\frac{\mathrm{Var}_T(x^{(j)})}{\mathrm{Var}_{\Tb}(x^{(j)})},
	\label{eq:initialKL2d}
\end{equation}
with $\mathrm{Var}_T(x^{(j)})=T/k^{(j)}(T)$, and each term is again nonnegative.
As in one dimension, $D_0$ decreases for $T<\Tb$ and increases for $T>\Tb$, at both baths, $\Tb=1$ and $\Tb=3$, over the whole range $0<T<6.6$ where the stiffnesses are positive (Fig.~\ref{fig:kl}(b)).
Equivalently, $D_0'(T)/(T-\Tb)>0$; evaluated analytically on a dense grid, this ratio stays well away from zero, with minima of about $0.41$ at $\Tb=1$ and $0.2$ at $\Tb=3$.
At $\Tb=3$, for example, the preparation at $T=1.1$, near the strong point $T_s\approx1.25$, starts farther away than the one at $T=2$ ($D_0\approx0.41$ versus $0.11$), but its slow coefficient is about half as large, and the two KL divergences cross at $t_\times\approx1.12$.

For the coupled example of \cref{app:avoided} the preparations are correlated Gaussians, and $D_0$ is the corresponding Gaussian divergence.
It is likewise monotonic on both sides of both baths, $\Tb=1$ and $\Tb=3.4$ (Fig.~\ref{fig:kl}(c)); at $\Tb=3.4$ the strong preparation starts three times farther away than the one at $T=2$ ($D_0\approx0.61$ versus $0.20$).

\paragraph*{Antiferromagnet.}
Here \cref{eq:kl-canonical} applies directly.
The macrostate distribution is $\pi(m,s;T)\propto\Omega(m,s)\,\e^{-E(m,s)/T}$, where $\Omega$ counts the microstates of the macrostate, and since all of them have the same energy, the divergence between macrostate distributions equals the microscopic one.
$D_0$ is therefore monotonic at every bath and every $N$.
For $N\to\infty$ it takes a simple form.
Each distribution concentrates on its free-energy minimum, $\ln\pi(m,s;T)=-\Brk{\mathcal F(m,s,T)-\mathcal F(\avg{m}_T,\avg{\abs{s}}_T,T)}/T+o(N)$, where $\avg{\abs{s}}_T$ is the staggered magnetization at the minimum, nonzero only for antiferromagnetic preparations (we use $\abs{s}$ because $\avg{s}_T=0$ by symmetry; the two symmetric minima give the same value).
We write the divergence as $D_0=\avg{\ln\pi(\,\cdot\,;T)}_T-\avg{\ln\pi(\,\cdot\,;\Tb)}_T$ and evaluate both averages at the preparation macrostate $(\avg{m}_T,\avg{\abs{s}}_T)$, on which $\pi(\,\cdot\,;T)$ is concentrated.
The first term is then the log-probability of the preparation at its own minimum, which by the expression above is only $o(N)$.
The divergence per spin therefore comes from the second term alone, minus the log-probability of the preparation macrostate in the bath distribution,
\begin{equation}
	\lim_{N\to\infty}\frac{D_0(T)}{N}
	=-\lim_{N\to\infty}\frac1N\ln\pi(\avg{m}_T,\avg{\abs{s}}_T;\Tb)
	=\frac{\mathcal F(\avg{m}_T,\avg{\abs{s}}_T,\Tb)-\mathcal F(\avg{m}_{\Tb},0,\Tb)}{N\Tb},
	\label{eq:kl-afm}
\end{equation}
which is nonnegative, since the bath free energy is smallest at the bath minimum.
The divergence per spin is thus the free-energy cost, in the bath, of the preparation macrostate, and like the coefficients it requires only equilibrium properties.
As a numerical confirmation, we evaluated \cref{eq:kl-afm} at every paramagnetic bath on every fourth field of the grid of Fig.~\ref{fig:phase-diagram} ($35334$ baths, $381$ preparations each), and it increases monotonically with $\abs{T-\Tb}$ at all of them; Fig.~\ref{fig:kl}(d) shows the two baths of Fig.~\ref{fig:two-d}.

\section{A coupled example with an avoided crossing}
\label{app:avoided}
\setcounter{figure}{0}

A cross term in the Hessian rotates the normal directions of the bath.
For two coordinates with unit mobility, as in \cref{eq:ou2d-potential} but with a coupling $c$,
\begin{equation}
\begin{aligned}
 K&=\begin{pmatrix}k^{(1)}&c\\c&k^{(2)}\end{pmatrix},\\
 \lambda^{(1,2)}&=\frac{k^{(1)}+k^{(2)}}{2}\mp
 \sqrt{\brk{\frac{k^{(1)}-k^{(2)}}{2}}^2+c^2}.
\end{aligned}
 \label{eq:twohessian}
\end{equation}
The rule must then be applied to the slow eigenvector of the \emph{bath} matrix $K(\Tb)$ rather than to either of the original coordinates.
A nonzero coupling at $k^{(1)}=k^{(2)}$ produces an avoided crossing with gap $2\abs{c}$; an exact crossing requires both $k^{(1)}=k^{(2)}$ and $c=0$, and at an exact degeneracy one must project onto the entire degenerate eigenspace.
For a given quench, $K(\Tb)$ fixes the slow direction, while the equilibrium preparations fix its coefficient.

\begin{figure*}[t]
\centering
\includegraphics[width=\textwidth]{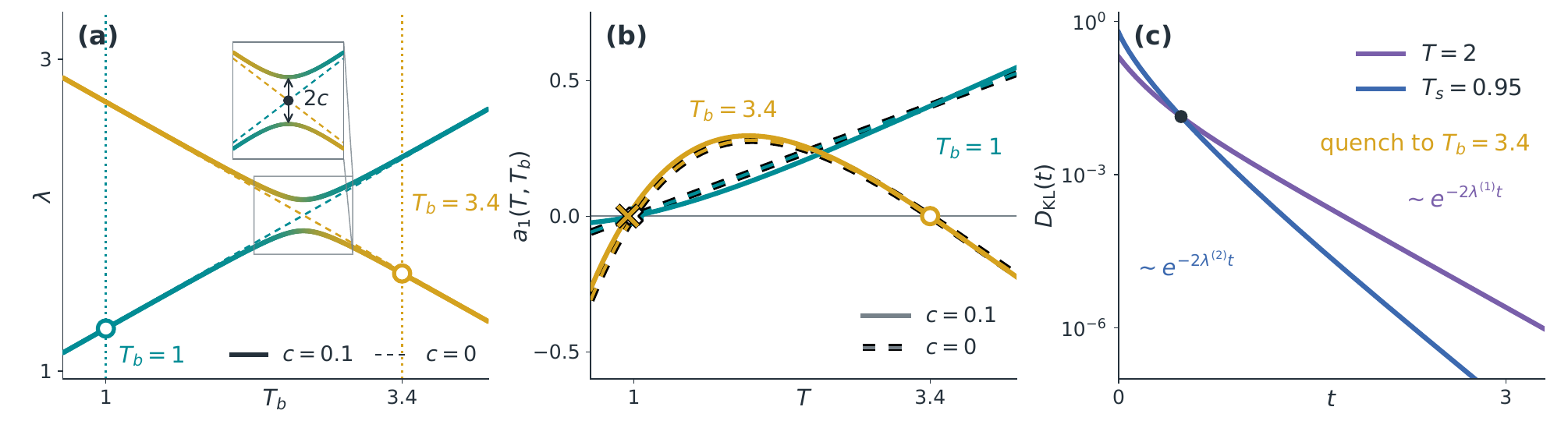}
	\caption{Coupled OU example, \cref{eq:avoided}, with $c=0.1$ (solid) and the uncoupled model $c=0$ (dashed), at a bath below the crossing, $\Tb=1$ (teal), and above it, $\Tb=3.4$ (gold).
	(a) Bath rates, \cref{eq:twohessian}: the coupling opens a gap $2c$ at $T=2.6$ (inset), where the rates of the uncoupled model cross (dot).
	The solid curves are colored by the direction of their eigenvector, from the $x^{(1)}$ axis (teal), along which the mean is monotonic, to the $x^{(2)}$ axis (gold), along which it is not; open circles mark the two baths.
	(b) Slow coefficient at the two baths; open circles mark the bath, and the cross the strong point.
	At $\Tb=1$ it follows the monotonic $\avg{x^{(1)}}_T$, with no effect; at $\Tb=3.4$ it follows the nonmonotonic $\avg{x^{(2)}}_T$ and vanishes at $T_s\approx0.95$ ($1.00$ for $c=0$).
	(c) KL divergence after quenches to $\Tb=3.4$ from the strong preparation $T_s$ and from $T=2$: the farther preparation overtakes the closer one (dot), an inverse Mpemba effect.}
\label{fig:avoided}
\end{figure*}

To show that coupling only rotates the rule, we take
\begin{equation}
 K(T)=
 \begin{pmatrix}
 2+0.45\brk{T-2.6}&c\\c&2-0.45\brk{T-2.6}
 \end{pmatrix},\qquad c=0.1,
 \label{eq:avoided}
\end{equation}
with the means
\begin{equation}
 \avg{x^{(1)}}_T=0.15\brk{T-1},\qquad
 \avg{x^{(2)}}_T=2\Brk{T\cdot3.4^{-\brk{T-1}/2.4}-1},
 \label{eq:avoided-mean}
\end{equation}
which have the same shapes as in \cref{eq:uncoupled2d}: $\avg{x^{(1)}}_T$ is monotonic, and $\avg{x^{(2)}}_T$ grows up to $T\approx2$ and then decreases.
By \cref{eq:twohessian} the two rates never cross: they approach each other at $T=2.6$, where the gap takes its minimum value, $2c=0.2$ (Fig.~\ref{fig:avoided}(a)).
Setting $c=0$ recovers an uncoupled model of the type of \cref{sec:two-dimensional}, whose rates, the two stiffnesses, cross exactly at $T=2.6$ (dashed in Fig.~\ref{fig:avoided}(a)).
We follow the slow coefficient at two baths, one on each side of the crossing, by projecting the preparation path on the slow eigenvector of $K(\Tb)$ (Fig.~\ref{fig:avoided}(b)).

\paragraph*{Below the crossing, $\Tb=1$.}
The bath rates are $\lambda^{(1)}\approx1.27$ and $\lambda^{(2)}\approx2.73$.
The coupling tilts the slow eigenvector of $K(\Tb)$ away from the $x^{(1)}$ axis, but only by about $4^\circ$, so $a_1$ is essentially the displacement of the monotonic mean $\avg{x^{(1)}}_T$.
It is monotonic in $T$, and there is no Mpemba effect, as for $c=0$.

\paragraph*{Above the crossing, $\Tb=3.4$.}
The bath rates are $\lambda^{(1)}\approx1.63$ and $\lambda^{(2)}\approx2.37$, and the slow eigenvector now lies within about $8^\circ$ of the $x^{(2)}$ axis, so $a_1$ follows the nonmonotonic mean $\avg{x^{(2)}}_T$.
It vanishes at the bath and again at $T_s\approx0.95<\Tb$, a strong inverse Mpemba effect, compared with $T_s=1.00$ for $c=0$: the avoided crossing only slightly shifts the strong point.
The preparation at $T_s$ starts three times farther from equilibrium than the one at $T=2$ ($D_0\approx0.61$ versus $0.20$), yet its KL divergence crosses the other at $t_\times\approx0.48$ and then decays at the faster rate (Fig.~\ref{fig:avoided}(c)).

As in \cref{sec:two-dimensional}, the slow moment switches from $\avg{x^{(1)}}$ to $\avg{x^{(2)}}$ across the crossing, and with it the existence of the effect; the coupling only rotates the slow direction slightly and turns the crossing into an avoided one.
The strong point therefore requires neither an exact eigenvalue crossing nor decoupled coordinates.

\section{Exact relaxation of the antiferromagnet for \texorpdfstring{$N\to\infty$}{N to infinity}}
\label{app:afm-ninf}
\setcounter{figure}{0}

Here we derive the equations of motion of $\avg m$ and $\avg{s^2}$, the exact rates, \cref{eq:afm-rates}, and the exact slow coefficients, \cref{eq:a1-limit}, of the antiferromagnet.
We follow the setup of Klich \emph{et al.}~\cite{Klich2019}.

\paragraph*{Sublattice magnetizations and Glauber rates.}
It is convenient to describe a macrostate by the magnetizations of the two sublattices, $x_1=m+s$ and $x_2=m-s$, that is, by the average spin on each sublattice, $x_a=\brk{N^{(a)}_\uparrow-N^{(a)}_\downarrow}/n$, with $n=N/2$ spins per sublattice.
In the paramagnetic phase both sublattices are equally magnetized, $x_1=x_2$ and $s=0$; in the antiferromagnetic phase they are not.
In these variables the energy of \cref{eq:ms} reads $E=n\,x_1x_2-nH\brk{x_1+x_2}$, so each spin interacts only with the average spin of the other sublattice.
A spin on sublattice $a$ therefore feels the effective field $H-x_{\bar a}$, where $\bar a$ denotes the other sublattice, and flipping it from $\sigma=\pm1$ to $-\sigma$ costs the energy $2\sigma\brk{H-x_{\bar a}}$.
In Glauber dynamics this spin flips at the rate
\begin{equation}
	w_a(\sigma)=\frac14\Brk{1-\sigma\,\tau\bigbrk{x_{\bar a}}},\qquad \tau(x)=\tanh\frac{H-x}{\Tb},
	\label{eq:glauber-rate}
\end{equation}
which obeys detailed balance with respect to the bath, $w_a(\sigma)/w_a(-\sigma)=\e^{-2\sigma\brk{H-x_{\bar a}}/\Tb}$; the prefactor sets the unit of time.
Since all spins of a sublattice are equivalent, the dynamics closes on the sublattice magnetizations $(x_1,x_2)$.
Sublattice $a$ has $n\brk{1-x_a}/2$ down spins and $n\brk{1+x_a}/2$ up spins, and summing \cref{eq:glauber-rate} over them gives the rates at which a spin flip on sublattice $a$ raises $x_a$ by $2/n$ or lowers it by $2/n$
\begin{equation}
 W_{a,+}=\frac{n\brk{1-x_a}}{8}\Brk{1+\tau\bigbrk{x_{\bar a}}},\qquad
 W_{a,-}=\frac{n\brk{1+x_a}}{8}\Brk{1-\tau\bigbrk{x_{\bar a}}}.
 \label{eq:macro-rates}
\end{equation}

\paragraph*{Fokker--Planck equation.}
Each flip changes $x_a$ by $\pm2/n$.
The mean and the mean square of these jumps per unit time, the first two jump moments, are
\begin{equation}
 \frac{2}{n}\brk{W_{a,+}-W_{a,-}}=\frac{\tau\bigbrk{x_{\bar a}}-x_a}{2},\qquad
 \frac{4}{n^2}\brk{W_{a,+}+W_{a,-}}=\frac{1-x_a\tau\bigbrk{x_{\bar a}}}{n}.
 \label{eq:jump-moments}
\end{equation}
The first is the deterministic drift of $x_a$, and the second, of order $1/n$, is the strength of its noise.
Truncating the Kramers--Moyal expansion of the master equation after these two terms gives the Fokker--Planck equation \cref{eq:glauber-fp}, written there in terms of $m$ and $s$.

\paragraph*{Deterministic flow.}
For $N\to\infty$ the noise vanishes, and a distribution concentrated on one macrostate stays concentrated and moves with the drift, the mean-field flow
\begin{equation}
\begin{aligned}
 \dot m&=\frac{\tau\bigbrk{m-s}+\tau\bigbrk{m+s}}{4}-\frac m2,\\
 \dot s&=\frac{\tau\bigbrk{m-s}-\tau\bigbrk{m+s}}{4}-\frac s2 .
\end{aligned}
 \label{eq:afm-flow}
\end{equation}
Exchanging the two sublattices, $x_1\leftrightarrow x_2$, maps $s\mapsto-s$ and leaves the energy and the rates unchanged.
The flow is therefore invariant under $s\mapsto-s$, and a distribution that is symmetric under the exchange, as every equilibrium preparation is, stays symmetric, with $\avg{s}=0$ at all times.

\paragraph*{Moment equations.}
For a paramagnetic preparation $m$ is sharp, while $s$ fluctuates around zero with a Gaussian distribution of variance $\avg{s^2}=O(1/N)$.
Since $\avg s=0$, the leading moment of $s$ is $\avg{s^2}$.
Multiplying \cref{eq:glauber-fp} by $s^2$ and integrating by parts gives
\begin{equation}
\begin{aligned}
 \frac{\dd\avg{s^2}}{\dd t}={}\avg{s\,\frac{\tau\bigbrk{m-s}-\tau\bigbrk{m+s}}{2}}-\avg{s^2}
 +\frac{\avg{2-x_1\tau\bigbrk{x_2}-x_2\tau\bigbrk{x_1}}}{2N}.
\end{aligned}
 \label{eq:s2-exact}
\end{equation}
For small $s$ we expand both averages to leading order,
\begin{equation}
	\frac{\tau\bigbrk{m-s}-\tau\bigbrk{m+s}}{2}\simeq q(m)\,s,\qquad
	2-x_1\tau\bigbrk{x_2}-x_2\tau\bigbrk{x_1}\simeq2\,\Gamma(m),
	\label{eq:q-gamma}
\end{equation}
with $q(m)=\Brk{1-\tau\bigbrk{m}^2}/\Tb$, the slope of $-\tau$, which sets how strongly a staggered deviation is restored, and $\Gamma(m)=1-m\,\tau\bigbrk{m}$, the strength of the noise.
In terms of $\sigma=N\avg{s^2}$ this gives
\begin{equation}
	\dot m=\frac{\tau\bigbrk{m}-m}{2},\qquad
	\dot\sigma=-\Brk{1-q(m)}\sigma+\Gamma(m),
	\label{eq:m-sigma}
\end{equation}
where the first equation is the $m$ component of \cref{eq:afm-flow} at $s=0$.

\paragraph*{Rates.}
Linearizing \cref{eq:m-sigma} around the bath, with $\delta m=\avg{m}-\avg{m}_{\Tb}$, $\delta\sigma=\sigma-\sigma_b$ and $\sigma_b=N\avg{s^2}_{\Tb}$, gives the triangular system
\begin{equation}
 \frac{\dd}{\dd t}\begin{pmatrix}\delta m\\ \delta\sigma\end{pmatrix}
 =-\begin{pmatrix}\lambda^{(m)}&0\\ -\brk{\Gamma'+q'\sigma_b}&2\lambda^{(s)}\end{pmatrix}
 \begin{pmatrix}\delta m\\ \delta\sigma\end{pmatrix},
 \label{eq:afm-linear}
\end{equation}
with derivatives taken at the bath: $\avg{m}$ relaxes on its own, while $\avg{s^2}$ is driven by $\avg{m}$ but does not drive it back.
The diagonal entries are the rates of \cref{eq:afm-rates}, $\lambda^{(m)}=\brk{1+q_b}/2$ and $2\lambda^{(s)}=1-q_b$, with $q_b=q(\avg{m}_{\Tb})=\brk{1-\avg{m}_{\Tb}^2}/\Tb$, in agreement with \cite{Klich2019}.
The factor $2$ reflects the relaxation of $s$ itself: near $s=0$ the $s$ component of \cref{eq:afm-flow} reads $\dot s=-\Brk{1-q(m)}s/2$, which at the bath is $\dot s=-\lambda^{(s)}s$, so $s^2$ relaxes at $2\lambda^{(s)}$.
These rates take $t\to\infty$ before $N\to\infty$, as in the spectral definition of the effect \cite{Klich2019}, so a symmetric preparation has no overlap with the mode odd in $s$, at rate $\lambda^{(s)}$.
In the opposite order, as in \cite{Blom2026}, $\abs{s}$ of an antiferromagnetic preparation relaxes at $\lambda^{(s)}<\lambda^{(m)}$, so the staggered mode is the slowest, even for a symmetric preparation.
These are also the harmonic rates of \cref{eq:stiffness}, and the OU process of \cref{eq:ou-fp} reproduces the equilibrium variances of $m$ and $s$ at the bath.

\paragraph*{Slow coefficient from the flow.}
The coefficients $a_n=\avg{u_n}_T$ of \cref{eq:a_n} are averages of the left eigenfunctions $u_n$, the eigenfunctions of the adjoint $\mathcal L^\dagger$ of the Fokker--Planck operator $\mathcal L_{\rm FP}$ of \cref{eq:master}, $\mathcal L^\dagger u_n=-\lambda_n u_n$.
For \cref{eq:glauber-fp},
\begin{equation}
	\mathcal L^\dagger=\dot m(m,s)\,\partial_m+\dot s(m,s)\,\partial_s+O(1/N),
	\label{eq:backward}
\end{equation}
where $\dot m(m,s)$ and $\dot s(m,s)$ denote the right-hand sides of \cref{eq:afm-flow}, functions of the macrostate $(m,s)$, and the $O(1/N)$ terms are the diffusion.
For eigenfunctions that vary on scales of order one the diffusion is negligible as $N\to\infty$, and the eigenvalue equation becomes first order,
\begin{equation}
	\dot m(m,s)\,\partial_m u+\dot s(m,s)\,\partial_s u=-\lambda u.
	\label{eq:eig-flow}
\end{equation}
Its left-hand side is the rate of change of $u$ along the flow.
Let $\bigbrk{m(t),s(t)}$ be the trajectory of \cref{eq:afm-flow} that starts at the macrostate $(m,s)$; $u$ remains a fixed function of the macrostate, evaluated at the point that the trajectory reaches at time $t$ (not an average).
By the chain rule and \cref{eq:eig-flow}, $\dd u\bigbrk{m(t),s(t)}/\dd t=-\lambda\,u\bigbrk{m(t),s(t)}$, so $u$ decays exponentially along every trajectory, and for every $t$
\begin{equation}
 u(m,s)=\e^{\lambda t}\,u\bigbrk{m(t),s(t)}.
 \label{eq:flow-eigenfunction}
\end{equation}
We normalize the eigenfunctions by their linear part at the bath, $u^{(m)}\simeq m-\avg{m}_{\Tb}$ and $u^{(s)}\simeq s$.
Since every trajectory ends at the bath, \cref{eq:flow-eigenfunction} determines $u$ everywhere from its form near the bath: we follow the flow from $(m,s)$ until it is close to the bath, where $u^{(m)}\simeq m(t)-\avg{m}_{\Tb}$, and multiply by $\e^{\lambda^{(m)}t}$.
Since each preparation is concentrated on a single macrostate, $a^{(m)}=u^{(m)}(\avg{m}_T,\avg{\abs{s}}_T)$, which is the late-time amplitude of $\avg{m}$ in \cref{eq:a1-limit}.
Since \cref{eq:eig-flow} is of first order, $\bigbrk{u^{(s)}}^2$ is also an eigenfunction, with rate $2\lambda^{(s)}$.
It carries the $s^2$ mode of antiferromagnetic preparations, where $\avg{s^2}_T=\avg{\abs{s}}_T^2$ is of order one.
Since it is positive, these preparations never produce a strong point in that mode.
For paramagnetic preparations, on the other hand, the $s^2$ mode lives on the scale $s\sim N^{-1/2}$, where the noise is not negligible, and we treat it through $\sigma=N\avg{s^2}$.

\paragraph*{Paramagnetic preparations: closed forms.}
A paramagnetic preparation starts on the line $s=0$, which the flow never leaves, and on it the flow is one-dimensional, $\dot m=\Brk{\tau\bigbrk{m}-m}/2$.
\cref{eq:eig-flow} then reduces to $\dot m\,\dd u^{(m)}/\dd m=-\lambda^{(m)}u^{(m)}$, that is, $\bigbrk{\ln\abs{u^{(m)}}}'=-\lambda^{(m)}/\dot m$.
Near the bath $\dot m\simeq-\lambda^{(m)}\brk{m-\avg{m}_{\Tb}}$, so the right-hand side behaves as $1/\brk{m-\avg{m}_{\Tb}}$, whose integral gives the linear normalization $u^{(m)}\simeq m-\avg{m}_{\Tb}$.
Separating this part,
\begin{equation}
 \bigbrk{\ln\abs{u^{(m)}}}'=\frac{1}{m-\avg{m}_{\Tb}}-\brk{\frac{\lambda^{(m)}}{\dot m}+\frac{1}{m-\avg{m}_{\Tb}}},
 \label{eq:am-step}
\end{equation}
where the bracket is regular at the bath, and integrating from the bath to $\avg{m}_T$ gives
\begin{equation}
 a^{(m)}(T,\Tb)=\brk{\avg{m}_T-\avg{m}_{\Tb}}
 \exp\!\Brk{-\int_{\avg{m}_{\Tb}}^{\avg{m}_T}\!\brk{\frac{\lambda^{(m)}}{\dot m}+\frac{1}{m-\avg{m}_{\Tb}}}\dd m} .
 \label{eq:am-exact}
\end{equation}
The prefactor is the harmonic coefficient, and the exponential is the correction from the nonlinear flow; it equals one when the flow is linear, $\dot m=-\lambda^{(m)}\brk{m-\avg{m}_{\Tb}}$.
The derivative of \cref{eq:am-exact} with respect to $\avg{m}_T$ is $-\lambda^{(m)}a^{(m)}/\dot m>0$, since the flow points back to the bath, so $\dot m$ and $a^{(m)}$ have opposite signs.
The exact $a^{(m)}$ is therefore a monotonic function of $\avg{m}_T$: it vanishes and turns back exactly where the moment displacement $\avg{m}_T-\avg{m}_{\Tb}$ does, and both give the same classification.

For the $s^2$ mode of a paramagnetic preparation we use $\sigma=N\avg{s^2}$, which obeys \cref{eq:m-sigma}, $\dot\sigma=\Gamma(m)-\Brk{1-q(m)}\sigma$.
Because $\dot\sigma$ is linear in $\sigma$, the eigenvalue equation $\dot m\,\partial_m u+\dot\sigma\,\partial_\sigma u=-\brk{1-q_b}u$, with rate $2\lambda^{(s)}=1-q_b$, has a solution linear in $\sigma$,
\begin{equation}
 u^{(s^2)}=g(m)\Brk{\sigma+w(m)},\qquad
 \brk{\ln g}'=-\frac{q-q_b}{\dot m},\qquad
 \dot m\,w'+\brk{1-q}w=-\Gamma,
 \label{eq:ass-exact}
\end{equation}
where the two equations for the functions $g(m)$ and $w(m)$ follow from collecting the terms with and without $\sigma$, and we take their solutions regular at the bath, with $g(\avg{m}_{\Tb})=1$ and $w(\avg{m}_{\Tb})=-\sigma_b$.
The exact coefficient is then $a^{(s^2)}(T,\Tb)=g(\avg{m}_T)\Brk{N\avg{s^2}_T+w(\avg{m}_T)}$, whereas the harmonic one, the moment displacement $N\brk{\avg{s^2}_T-\avg{s^2}_{\Tb}}$, corresponds to $g=1$ and $w=-\sigma_b$.
The difference comes mainly from $w$, which describes the drive of $\avg{s^2}$ by $\avg{m}$.
Its slope at the bath,
\begin{equation}
 w'(\avg{m}_{\Tb})=\frac{2\brk{\Gamma'+q'\sigma_b}}{3q_b-1}\bigg|_{\avg{m}_{\Tb}},
 \label{eq:resonance}
\end{equation}
diverges at the rate crossing, $q_b=1/3$, where $\avg m$ and $\avg{s^2}$ relax at the same rate and this drive is resonant.
This is why the two phase diagrams of Fig.~\ref{fig:phase-diagram} differ most where $s^2$ is slow.

\paragraph*{Antiferromagnetic preparations.}
These preparations lie off the line $s=0$, where the flow is two-dimensional and $a^{(m)}$ has no closed form, so we evaluate \cref{eq:flow-eigenfunction} numerically.
We integrate the flow, \cref{eq:afm-flow}, from the preparation until both sublattice magnetizations are within $10^{-6}$ of $\avg{m}_{\Tb}$, at some time $t$.
There we evaluate $u^{(m)}$ by its second-order expansion around the bath and multiply it by $\e^{\lambda^{(m)}t}$; the error is of third order in the remaining deviation.

\section{Equilibrium moments, exact dynamics and classification for the antiferromagnet}
\label{app:afm-details}
\setcounter{figure}{0}

This appendix collects the equilibrium input of the phase diagrams and the rate crossing.

\paragraph*{Equilibrium preparations.}
All coefficients, and both phase diagrams, use the large-$N$ free energy $\mathcal F(m,s,T)$ of \cref{sec:antiferromagnet}, and each preparation is taken at its global minimum.
For $T>1-m^2$ the minimum is paramagnetic, at $s=0$ with $m=\tanh\Brk{\brk{H-m}/T}$.
Below this line it is antiferromagnetic, with $x_a=\tanh\Brk{\brk{H-x_{\bar a}}/T}$ and $x_1\neq x_2$, and the transition between the two is continuous.
The staggered moment $\avg{s^2}_T$ is of a different order in the two phases:
\begin{itemize}
	\item In the paramagnetic phase $s$ only fluctuates around zero, and $\avg{s^2}_T$ is of order $1/N$, $N\avg{s^2}_T=\brk{1-m^2}/\brk{1-q_T}$ with $q_T=\brk{1-m^2}/T$, the stationary value of \cref{eq:m-sigma} at temperature $T$. Equivalently, $N\avg{s^2}_T=T\chi_s$, with $\chi_s$ the staggered susceptibility, the quantity used as a slow-mode proxy in \cite{GonzalezAdalid2024,Pomares2026}, which diverges at the transition, $q_T=1$.
	\item In the antiferromagnetic phase $\avg{s^2}_T=\avg{\abs{s}}_T^2$ is of order one, with $\avg{\abs{s}}_T$ the staggered magnetization at the minimum.
\end{itemize}
The two figures of the main text use these values at two different orders in $1/N$.
\Cref{fig:two-d}(d) shows $\avg{s^2}_T$ itself for $N\to\infty$: it vanishes for paramagnetic preparations, equals $\avg{\abs{s}}_T^2$ for antiferromagnetic ones, and is continuous across the transition.
\Cref{fig:two-d}(f) and the $s^2$ coefficient, \cref{eq:ass-exact}, use instead $N\avg{s^2}_T$, which resolves the paramagnetic fluctuations; in these units antiferromagnetic preparations, with $\avg{s^2}_T$ of order one, are off the scale and are drawn at the top of the rescaled axis.

\paragraph*{Rate crossing.} 
The rates of \cref{eq:afm-rates}, which are also the harmonic rates of \cref{eq:stiffness}, exchange where $\lambda^{(m)}=2\lambda^{(s)}$, that is at $q_b=1/3$.
Parametrized by the bath magnetization, this is the curve
\begin{equation}
 T_\times=3\brk{1-\avg{m}_{\Tb}^2},\qquad
 H_\times=\avg{m}_{\Tb}+T_\times\operatorname{artanh}\avg{m}_{\Tb},
 \label{eq:crosscurve}
\end{equation}
which reaches $\Tb=3$ as $H\to0^+$, and is the dashed line in Fig.~\ref{fig:phase-diagram}.

\section{Where the exact and harmonic phase diagrams differ}
\label{app:disagreement}
\setcounter{figure}{0}

\begin{figure*}[t]
	\centering
	\includegraphics[width=\textwidth]{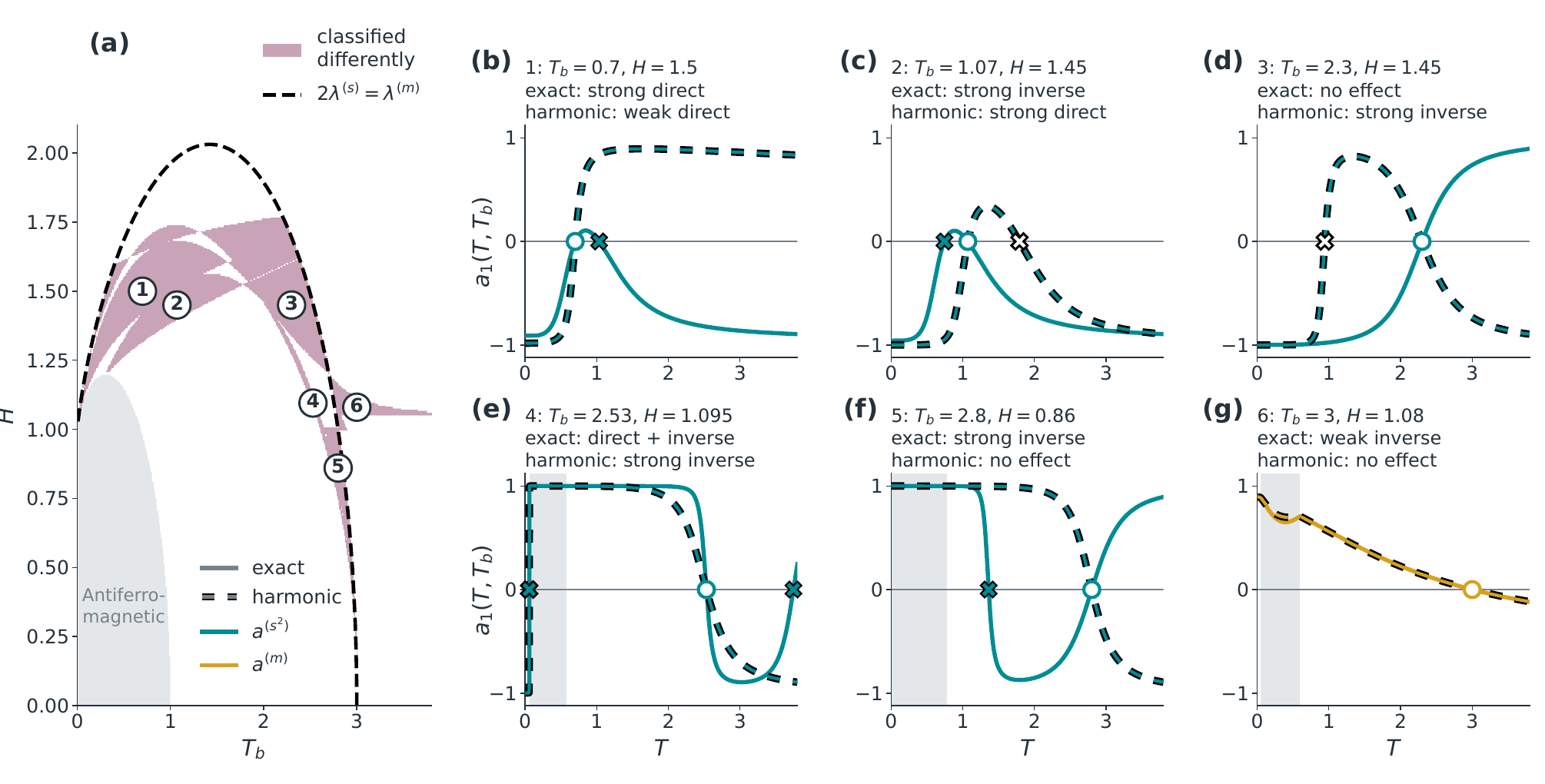}
	\caption{Baths at which the exact and harmonic phase diagrams of Fig.~\ref{fig:phase-diagram} differ.
		(a) Paramagnetic baths classified differently (shaded), with the rate crossing $2\lambda^{(s)}=\lambda^{(m)}$ (dashed) and the six baths of panels (b)--(g) (numbered).
		(b)--(g) Slow coefficient $a_1(T,\Tb)$ along the preparations at these baths, exact (solid) and harmonic (dashed), with the classification of each written above the panel.
		Teal curves are the $s^2$ coefficient and gold the $m$ coefficient; each curve is rescaled as in Fig.~\ref{fig:two-d}(f).
		Open circles mark the bath, crosses the strong points (filled: exact; open: harmonic), and gray bands the antiferromagnetic preparations.}
	\label{fig:disagreement}
\end{figure*}

Here we focus on the baths at which the exact and harmonic classifications differ, to understand where and why the harmonic approximation fails.
Over most of the phase diagram the two agree: of the $140005$ paramagnetic baths of Fig.~\ref{fig:phase-diagram}, $88.2\%$ are classified identically, and for $H<1$ the agreement reaches $97.9\%$, with the two diagrams almost indistinguishable.
The remaining $11.8\%$ lie mostly at $H>1$, where they make up $18.7\%$ of the baths, and $93.6\%$ of them are baths at which $s^2$ is slow; most of the others, at which $m$ is slow, lie in the thin weak-inverse band just above $H=1$.
Figure~\ref{fig:disagreement}(a) shows where the two classifications differ, and panels (b)--(g) compare the two coefficients at six representative baths.

Where $s^2$ is slow, the two coefficients differ through the drive of $\avg{s^2}$ by $\avg m$, which the harmonic coefficient omits ($g=1$ and $w=-\sigma_b$ in \cref{eq:ass-exact}).
This drive matters in two regimes.
Near the crossing it is resonant, \cref{eq:resonance}, and large even for preparations close to the bath (baths 3--5): it can create a strong point (bath 5, the narrow strip along the crossing in Fig.~\ref{fig:phase-diagram}(a)), remove one (bath 3), or add one on the other side of the bath (bath 4).
Far from the crossing it is not resonant, but for $H>1$ the preparations span a wide range of $\avg{m}_T$, so $g$ and $w$ are evaluated far from the bath; this can turn a weak effect into a strong one (bath 1), or move the strong point to the other side of the bath and reverse its direction (bath 2).
Where $m$ is slow (bath 6), the two coefficients are almost indistinguishable, and they differ only through the re-entrant window described next.

\paragraph*{Re-entrant window near $H=1$.} 
Just above $H=1$ the antiferromagnetic phase is re-entrant: it occupies a window of intermediate temperatures, below which the equilibrium state is again fully polarized (gray in Fig.~\ref{fig:phase-diagram}).
A re-entrant transition is known to produce a strong Mpemba effect through the staggered order \cite{Blom2026}, which in our diagram corresponds to the baths where $s^2$ is slow (e.g.\ the strong inverse effect at $H=1.1$); here, where $m$ is slow, it acts through $\avg{m}_T$.
Inside this window the staggered order lowers $\avg{m}_T$, and this dip makes $\abs{a^{(m)}}$ nonmonotonic for cold preparations, which produces the thin weak-inverse band where $m$ is slow.
The dip is shallow in the harmonic coefficient and is deepened by the nonlinear flow in the exact one (Fig.~\ref{fig:disagreement}(g)), so the band is wider in the exact diagram; being set by a shallow dip, its width is sensitive to the weak-effect threshold.
The re-entrance itself is a mean-field feature, absent on the square lattice \cite{Blom2026}.

In all cases the slow moment, selected by the exact rates, is the same for both coefficients.
The two diagrams differ in the strength or the direction of the effect, not in which moment controls it.

\bibliography{references}

\end{document}